\documentclass[a4paper,11pt,titlepage,oneside]{book} %it was scr book
\usepackage{natbib}
\usepackage{pdflscape}
\usepackage{epstopdf}
\usepackage{amsmath,amssymb}
\usepackage[pdfborder={0 0 0}]{hyperref}
\hypersetup{pdfinfo={
  Title={Performance-oriented DevOps: A Research Agenda},
  Author={Andreas Brunnert, Andre van Hoorn, Felix Willnecker, Alexandru Danciu, Wilhelm Hasselbring, Christoph Heger, Nikolas Herbst, Pooyan Jamshidi, Reiner Jung, Joakim von Kistowski, Anne Koziolek, Johannes Kro{\ss}, Simon Spinner, Christian V{\"o}gele, J{\"u}rgen Walter, Alexander Wert},
  Subject={Technical Report: SPEC-RG-2015-01},
  Keywords={DevOps; Software Performance Engineering; Application Performance Management}}
}
\usepackage[]{acronym}
\usepackage[export]{adjustbox}

\usepackage[T1]{fontenc}
\usepackage[utf8]{inputenc}

\usepackage{ae}               % Almost european, virtual T1-Font
\usepackage[]{graphicx}
\usepackage{vmargin}          % Adjust margins in a simple way
\usepackage{fancyhdr}         % Define simple headings
\usepackage[font=footnotesize]{subfig}
\usepackage{pdfcomment}
\usepackage[absolute,overlay]{textpos}
\usepackage{tikz}
\usepackage[english]{babel}
\usepackage{color}
\usepackage{longtable}
\usepackage{hyperref} % Needs to be before glossaries to enable hyperlinks for acronyms
\usepackage[toc,acronym,nonumberlist]{glossaries} %nopostdot, additional list of abbreveations
\usepackage[bottom]{footmisc}
\usepackage{multirow}
\usepackage{array}
\usepackage{nomencl}
\usepackage{color}
\usepackage{xcolor}
\usepackage{setspace}

\makenomenclature

\makeglossary

\definecolor{darkred}{rgb}{0.5,0,0}
\definecolor{darkgreen}{rgb}{0,0.5,0}
\definecolor{darkblue}{rgb}{0,0,0.5}
\setmarginsrb
{2.5cm}   % left margin
{1.5cm}   % top margin
{2.5cm}   % right margin
{2cm}   % bottom margin
{20pt}    % head height
{0.25in}  % head sep
{9pt}     % foot height
{0.3in}   % foot sep

\newcommand{\summary}[1]{}

\usepackage{chngcntr}
\counterwithin{figure}{section}
\counterwithin{table}{section}

\newcommand{\forget}[1]{}  % text intentionally left out for whatever reason
\newcommand{\sandbox}[1]{}  % SANDBOX: internal multi-line comments by the author(s): reminders, notes, todos, detailed text blocks from related papers
\newcommand{\longer}[1]{}  % commented out text that is candidate for integrating in extended version of the paper / proposal

\newcommand{\shorten}[1]{}  % text currently shortened, but immediate candidate to put back in later if space permits ("ungerne gekürzt"), normally it should be possible to put the text back in by simply removing the tag without further changes needed
\usepackage[olditem,oldenum]{paralist} % for inline lists
\newenvironment{inlineenumerate}%
{%BEGIN:
\begin{inparaenum}[\itshape a\upshape)]
}%
{%END:
\end{inparaenum}
}

\usepackage{authblk}

\author[1]{Tentative author list\footnote{SPEC Confidential. This document should be considered confidential unless labeled otherwise.}}
\newcommand\mytitle{Performance-oriented DevOps: A Research Agenda}
\title{Performance-oriented DevOps: A Research Agenda}

\newcommand\TRnumber{%
Technical Report: SPEC-RG-2015-01%\\Version: 0.1
}
\newcommand\WGname{SPEC RG DevOps Performance Working Group}

\newcommand\TRdate{\today}
\newcommand\TRcentralURL{research.spec.org}
\newcommand\TRrightURL{www.spec.org}

\newcommand\Acknowledgements{
This work has been supported by the Research Group of the Standard %
Performance\linebreak Evaluation Corporation (SPEC), by %
the German Federal Ministry of Education and Research (Andr\'e van Hoorn, grant no.\ 01IS15004 (diagnoseIT)), and %
by the European Commission\linebreak (Pooyan Jamshidi, grant no.\ FP7-ICT-2011-8-318484 (MODAClouds) and H2020-ICT-2014-1-644869 (DICE)). %
}

\newcommand\numAuthors{1} %this amount of authors will be displayed
\makeatletter
\newcommand\defcase[1]{\@namedef{mycase@\the\numexpr#1\relax}}
\newcommand\putAuthors[1]{\@nameuse{mycase@\the\numexpr#1\relax}}
\makeatother

\defcase{0}{
	\begin{textblock}{8}[0,0](\colsinglecentralX,\rowoneY)
		\centering
		\small{\textbf{No Author}}\\
	\end{textblock}	
}

\defcase{1}{
	\begin{textblock}{8}[0,0](\colsinglecentralX,\rowoneY)
		\centering
		\small{\textbf{\authorOneName}}\\
		\footnotesize
		\authorOneAffil
	\end{textblock}	
}
\defcase{2}{
	\begin{textblock}{\authorCellWidth}[0,0](\colDoubleLeftX,\rowoneY)
		\centering
		\small{\textbf{\authorOneName}}\\
		\footnotesize
		\authorOneAffil
	\end{textblock}	
	\begin{textblock}{\authorCellWidth}[0,0](\colDoubleRightX,\rowoneY)
		\centering
		\small{\textbf{\authorTwoName}}\\
		\footnotesize
		\authorTwoAffil
	\end{textblock}	
}
\defcase{3}{
	\begin{textblock}{\authorCellWidth}[0,0](\coloneX,\rowoneY)
		\centering
		\small{\textbf{\authorOneName}}\\
		\footnotesize
		\authorOneAffil
	\end{textblock}	
	\begin{textblock}{\authorCellWidth}[0,0](\coltwoX,\rowoneY)
		\centering
		\small{\textbf{\authorTwoName}}\\
		\footnotesize
		\authorTwoAffil
	\end{textblock}	
	\begin{textblock}{\authorCellWidth}[0,0](\colthreeX,\rowoneY)
		\centering
		\small{\textbf{\authorThreeName}}\\
		\footnotesize
		\authorThreeAffil
	\end{textblock}	
}
\defcase{4}{
	\begin{textblock}{\authorCellWidth}[0,0](\coloneX,\rowoneY)
		\centering
		\small{\textbf{\authorOneName}}\\
		\footnotesize
		\authorOneAffil
	\end{textblock}	
	\begin{textblock}{\authorCellWidth}[0,0](\coltwoX,\rowoneY)
		\centering
		\small{\textbf{\authorTwoName}}\\
		\footnotesize
		\authorTwoAffil
	\end{textblock}	
	\begin{textblock}{\authorCellWidth}[0,0](\colthreeX,\rowoneY)
		\centering
		\small{\textbf{\authorThreeName}}\\
		\footnotesize
		\authorThreeAffil
	\end{textblock}
	\begin{textblock}{8}[0,0](\colsinglecentralX,\rowtwoY)
		\centering
		\small{\textbf{\authorFourName}}\\
		\footnotesize
		\authorFourAffil
	\end{textblock}	
}
\defcase{5}{
	\begin{textblock}{\authorCellWidth}[0,0](\coloneX,\rowoneY)
		\centering
		\small{\textbf{\authorOneName}}\\
		\footnotesize
		\authorOneAffil
	\end{textblock}	
	\begin{textblock}{\authorCellWidth}[0,0](\coltwoX,\rowoneY)
		\centering
		\small{\textbf{\authorTwoName}}\\
		\footnotesize
		\authorTwoAffil
	\end{textblock}	
	\begin{textblock}{\authorCellWidth}[0,0](\colthreeX,\rowoneY)
		\centering
		\small{\textbf{\authorThreeName}}\\
		\footnotesize
		\authorThreeAffil
	\end{textblock}
	\begin{textblock}{\authorCellWidth}[0,0](\colDoubleLeftX,\rowtwoY)
		\centering
		\small{\textbf{\authorFourName}}\\
		\footnotesize
		\authorFourAffil
	\end{textblock}	
	\begin{textblock}{\authorCellWidth}[0,0](\colDoubleRightX,\rowtwoY)
		\centering
		\small{\textbf{\authorFiveName}}\\
		\footnotesize
		\authorFiveAffil
	\end{textblock}	
}
\defcase{6}{
	\begin{textblock}{\authorCellWidth}[0,0](\coloneX,\rowoneY)
		\centering
		\small{\textbf{\authorOneName}}\\
		\footnotesize
		\authorOneAffil
	\end{textblock}	
	\begin{textblock}{\authorCellWidth}[0,0](\coltwoX,\rowoneY)
		\centering
		\small{\textbf{\authorTwoName}}\\
		\footnotesize
		\authorTwoAffil
	\end{textblock}	
	\begin{textblock}{\authorCellWidth}[0,0](\colthreeX,\rowoneY)
		\centering
		\small{\textbf{\authorThreeName}}\\
		\footnotesize
		\authorThreeAffil
	\end{textblock}
	\begin{textblock}{\authorCellWidth}[0,0](\coloneX,\rowtwoY)
		\centering
		\small{\textbf{\authorFourName}}\\
		\footnotesize
		\authorFourAffil
	\end{textblock}	
	\begin{textblock}{\authorCellWidth}[0,0](\coltwoX,\rowtwoY)
		\centering
		\small{\textbf{\authorFiveName}}\\
		\footnotesize
		\authorFiveAffil
	\end{textblock}	
	\begin{textblock}{\authorCellWidth}[0,0](\colthreeX,\rowtwoY)
		\centering
		\small{\textbf{\authorSixName}}\\
		\footnotesize
		\authorSixAffil
	\end{textblock}
}
\defcase{7}{
	\begin{textblock}{\authorCellWidth}[0,0](\coloneX,\rowoneY)
		\centering
		\small{\textbf{\authorOneName}}\\
		\footnotesize
		\authorOneAffil
	\end{textblock}	
	\begin{textblock}{\authorCellWidth}[0,0](\coltwoX,\rowoneY)
		\centering
		\small{\textbf{\authorTwoName}}\\
		\footnotesize
		\authorTwoAffil
	\end{textblock}	
	\begin{textblock}{\authorCellWidth}[0,0](\colthreeX,\rowoneY)
		\centering
		\small{\textbf{\authorThreeName}}\\
		\footnotesize
		\authorThreeAffil
	\end{textblock}
	\begin{textblock}{\authorCellWidth}[0,0](\coloneX,\rowtwoY)
		\centering
		\small{\textbf{\authorFourName}}\\
		\footnotesize
		\authorFourAffil
	\end{textblock}	
	\begin{textblock}{\authorCellWidth}[0,0](\coltwoX,\rowtwoY)
		\centering
		\small{\textbf{\authorFiveName}}\\
		\footnotesize
		\authorFiveAffil
	\end{textblock}	
	\begin{textblock}{\authorCellWidth}[0,0](\colthreeX,\rowtwoY)
		\centering
		\small{\textbf{\authorSixName}}\\
		\footnotesize
		\authorSixAffil
	\end{textblock}
	\begin{textblock}{8}[0,0](\colsinglecentralX,\rowthreeY)
		\centering
		\small{\textbf{\authorSevenName}}\\
		\footnotesize
		\authorSevenAffil
	\end{textblock}
}
\defcase{8}{
	\begin{textblock}{\authorCellWidth}[0,0](\coloneX,\rowoneY)
		\centering
		\small{\textbf{\authorOneName}}\\
		\footnotesize
		\authorOneAffil
	\end{textblock}	
	\begin{textblock}{\authorCellWidth}[0,0](\coltwoX,\rowoneY)
		\centering
		\small{\textbf{\authorTwoName}}\\
		\footnotesize
		\authorTwoAffil
	\end{textblock}	
	\begin{textblock}{\authorCellWidth}[0,0](\colthreeX,\rowoneY)
		\centering
		\small{\textbf{\authorThreeName}}\\
		\footnotesize
		\authorThreeAffil
	\end{textblock}
	\begin{textblock}{\authorCellWidth}[0,0](\coloneX,\rowtwoY)
		\centering
		\small{\textbf{\authorFourName}}\\
		\footnotesize
		\authorFourAffil
	\end{textblock}	
	\begin{textblock}{\authorCellWidth}[0,0](\coltwoX,\rowtwoY)
		\centering
		\small{\textbf{\authorFiveName}}\\
		\footnotesize
		\authorFiveAffil
	\end{textblock}	
	\begin{textblock}{\authorCellWidth}[0,0](\colthreeX,\rowtwoY)
		\centering
		\small{\textbf{\authorSixName}}\\
		\footnotesize
		\authorSixAffil
	\end{textblock}
	\begin{textblock}{\authorCellWidth}[0,0](\colDoubleLeftX,\rowthreeY)
		\centering
		\small{\textbf{\authorSevenName}}\\
		\footnotesize
		\authorSevenAffil
	\end{textblock}
	\begin{textblock}{\authorCellWidth}[0,0](\colDoubleRightX,\rowthreeY)
		\centering
		\small{\textbf{\authorEightName}}\\
		\footnotesize
		\authorEightAffil
	\end{textblock}
}
\defcase{9}{
	\begin{textblock}{\authorCellWidth}[0,0](\coloneX,\rowoneY)
		\centering
		\small{\textbf{\authorOneName}}\\
		\footnotesize
		\authorOneAffil
	\end{textblock}	
	\begin{textblock}{\authorCellWidth}[0,0](\coltwoX,\rowoneY)
		\centering
		\small{\textbf{\authorTwoName}}\\
		\footnotesize
		\authorTwoAffil
	\end{textblock}	
	\begin{textblock}{\authorCellWidth}[0,0](\colthreeX,\rowoneY)
		\centering
		\small{\textbf{\authorThreeName}}\\
		\footnotesize
		\authorThreeAffil
	\end{textblock}
	\begin{textblock}{\authorCellWidth}[0,0](\coloneX,\rowtwoY)
		\centering
		\small{\textbf{\authorFourName}}\\
		\footnotesize
		\authorFourAffil
	\end{textblock}	
	\begin{textblock}{\authorCellWidth}[0,0](\coltwoX,\rowtwoY)
		\centering
		\small{\textbf{\authorFiveName}}\\
		\footnotesize
		\authorFiveAffil
	\end{textblock}	
	\begin{textblock}{\authorCellWidth}[0,0](\colthreeX,\rowtwoY)
		\centering
		\small{\textbf{\authorSixName}}\\
		\footnotesize
		\authorSixAffil
	\end{textblock}
	\begin{textblock}{\authorCellWidth}[0,0](\coloneX,\rowthreeY)
		\centering
		\small{\textbf{\authorSevenName}}\\
		\footnotesize
		\authorSevenAffil
	\end{textblock}	
	\begin{textblock}{\authorCellWidth}[0,0](\coltwoX,\rowthreeY)
		\centering
		\small{\textbf{\authorEightName}}\\
		\footnotesize
		\authorEightAffil
	\end{textblock}	
	\begin{textblock}{\authorCellWidth}[0,0](\colthreeX,\rowthreeY)
		\centering
		\small{\textbf{\authorNineName}}\\
		\footnotesize
		\authorNineAffil
	\end{textblock}
}
\defcase{10}{
	\begin{textblock}{\authorCellWidth}[0,0](\coloneX,\rowoneY)
		\centering
		\small{\textbf{\authorOneName}}\\
		\footnotesize
		\authorOneAffil
	\end{textblock}	
	\begin{textblock}{\authorCellWidth}[0,0](\coltwoX,\rowoneY)
		\centering
		\small{\textbf{\authorTwoName}}\\
		\footnotesize
		\authorTwoAffil
	\end{textblock}	
	\begin{textblock}{\authorCellWidth}[0,0](\colthreeX,\rowoneY)
		\centering
		\small{\textbf{\authorThreeName}}\\
		\footnotesize
		\authorThreeAffil
	\end{textblock}
	\begin{textblock}{\authorCellWidth}[0,0](\coloneX,\rowtwoY)
		\centering
		\small{\textbf{\authorFourName}}\\
		\footnotesize
		\authorFourAffil
	\end{textblock}	
	\begin{textblock}{\authorCellWidth}[0,0](\coltwoX,\rowtwoY)
		\centering
		\small{\textbf{\authorFiveName}}\\
		\footnotesize
		\authorFiveAffil
	\end{textblock}	
	\begin{textblock}{\authorCellWidth}[0,0](\colthreeX,\rowtwoY)
		\centering
		\small{\textbf{\authorSixName}}\\
		\footnotesize
		\authorSixAffil
	\end{textblock}
	\begin{textblock}{\authorCellWidth}[0,0](\coloneX,\rowthreeY)
		\centering
		\small{\textbf{\authorSevenName}}\\
		\footnotesize
		\authorSevenAffil
	\end{textblock}	
	\begin{textblock}{\authorCellWidth}[0,0](\coltwoX,\rowthreeY)
		\centering
		\small{\textbf{\authorEightName}}\\
		\footnotesize
		\authorEightAffil
	\end{textblock}	
	\begin{textblock}{\authorCellWidth}[0,0](\colthreeX,\rowthreeY)
		\centering
		\small{\textbf{\authorNineName}}\\
		\footnotesize
		\authorNineAffil
	\end{textblock}
	\begin{textblock}{8}[0,0](\colsinglecentralX,\rowfourY)
		\centering
		\small{\textbf{\authorTenName}}\\
		\footnotesize
		\authorTenAffil
	\end{textblock}
}
\defcase{11}{
	\begin{textblock}{\authorCellWidth}[0,0](\coloneX,\rowoneY)
		\centering
		\small{\textbf{\authorOneName}}\\
		\footnotesize
		\authorOneAffil
	\end{textblock}	
	\begin{textblock}{\authorCellWidth}[0,0](\coltwoX,\rowoneY)
		\centering
		\small{\textbf{\authorTwoName}}\\
		\footnotesize
		\authorTwoAffil
	\end{textblock}	
	\begin{textblock}{\authorCellWidth}[0,0](\colthreeX,\rowoneY)
		\centering
		\small{\textbf{\authorThreeName}}\\
		\footnotesize
		\authorThreeAffil
	\end{textblock}
	\begin{textblock}{\authorCellWidth}[0,0](\coloneX,\rowtwoY)
		\centering
		\small{\textbf{\authorFourName}}\\
		\footnotesize
		\authorFourAffil
	\end{textblock}	
	\begin{textblock}{\authorCellWidth}[0,0](\coltwoX,\rowtwoY)
		\centering
		\small{\textbf{\authorFiveName}}\\
		\footnotesize
		\authorFiveAffil
	\end{textblock}	
	\begin{textblock}{\authorCellWidth}[0,0](\colthreeX,\rowtwoY)
		\centering
		\small{\textbf{\authorSixName}}\\
		\footnotesize
		\authorSixAffil
	\end{textblock}
	\begin{textblock}{\authorCellWidth}[0,0](\coloneX,\rowthreeY)
		\centering
		\small{\textbf{\authorSevenName}}\\
		\footnotesize
		\authorSevenAffil
	\end{textblock}	
	\begin{textblock}{\authorCellWidth}[0,0](\coltwoX,\rowthreeY)
		\centering
		\small{\textbf{\authorEightName}}\\
		\footnotesize
		\authorEightAffil
	\end{textblock}	
	\begin{textblock}{\authorCellWidth}[0,0](\colthreeX,\rowthreeY)
		\centering
		\small{\textbf{\authorNineName}}\\
		\footnotesize
		\authorNineAffil
	\end{textblock}
	\begin{textblock}{\authorCellWidth}[0,0](\colDoubleLeftX,\rowfourY)
		\centering
		\small{\textbf{\authorTenName}}\\
		\footnotesize
		\authorTenAffil
	\end{textblock}
	\begin{textblock}{\authorCellWidth}[0,0](\colDoubleRightX,\rowfourY)
		\centering
		\small{\textbf{\authorElevenName}}\\
		\footnotesize
		\authorElevenAffil
	\end{textblock}
}
\defcase{12}{
	\begin{textblock}{\authorCellWidth}[0,0](\coloneX,\rowoneY)
		\centering
		\small{\textbf{\authorOneName}}\\
		\footnotesize
		\authorOneAffil
	\end{textblock}	
	\begin{textblock}{\authorCellWidth}[0,0](\coltwoX,\rowoneY)
		\centering
		\small{\textbf{\authorTwoName}}\\
		\footnotesize
		\authorTwoAffil
	\end{textblock}	
	\begin{textblock}{\authorCellWidth}[0,0](\colthreeX,\rowoneY)
		\centering
		\small{\textbf{\authorThreeName}}\\
		\footnotesize
		\authorThreeAffil
	\end{textblock}
	\begin{textblock}{\authorCellWidth}[0,0](\coloneX,\rowtwoY)
		\centering
		\small{\textbf{\authorFourName}}\\
		\footnotesize
		\authorFourAffil
	\end{textblock}	
	\begin{textblock}{\authorCellWidth}[0,0](\coltwoX,\rowtwoY)
		\centering
		\small{\textbf{\authorFiveName}}\\
		\footnotesize
		\authorFiveAffil
	\end{textblock}	
	\begin{textblock}{\authorCellWidth}[0,0](\colthreeX,\rowtwoY)
		\centering
		\small{\textbf{\authorSixName}}\\
		\footnotesize
		\authorSixAffil
	\end{textblock}
	\begin{textblock}{\authorCellWidth}[0,0](\coloneX,\rowthreeY)
		\centering
		\small{\textbf{\authorSevenName}}\\
		\footnotesize
		\authorSevenAffil
	\end{textblock}	
	\begin{textblock}{\authorCellWidth}[0,0](\coltwoX,\rowthreeY)
		\centering
		\small{\textbf{\authorEightName}}\\
		\footnotesize
		\authorEightAffil
	\end{textblock}	
	\begin{textblock}{\authorCellWidth}[0,0](\colthreeX,\rowthreeY)
		\centering
		\small{\textbf{\authorNineName}}\\
		\footnotesize
		\authorNineAffil
	\end{textblock}
	\begin{textblock}{\authorCellWidth}[0,0](\coloneX,\rowfourY)
		\centering
		\small{\textbf{\authorTenName}}\\
		\footnotesize
		\authorTenAffil
	\end{textblock}	
	\begin{textblock}{\authorCellWidth}[0,0](\coltwoX,\rowfourY)
		\centering
		\small{\textbf{\authorElevenName}}\\
		\footnotesize
		\authorElevenAffil
	\end{textblock}	
	\begin{textblock}{\authorCellWidth}[0,0](\colthreeX,\rowfourY)
		\centering
		\small{\textbf{\authorTwelveName}}\\
		\footnotesize
		\authorTwelveAffil
	\end{textblock}
}
\defcase{11}{
	\begin{textblock}{\authorCellWidth}[0,0](\coloneX,\rowoneY)
		\centering
		\small{\textbf{\authorOneName}}\\
		\footnotesize
		\authorOneAffil
	\end{textblock}	
	\begin{textblock}{\authorCellWidth}[0,0](\coltwoX,\rowoneY)
		\centering
		\small{\textbf{\authorTwoName}}\\
		\footnotesize
		\authorTwoAffil
	\end{textblock}	
	\begin{textblock}{\authorCellWidth}[0,0](\colthreeX,\rowoneY)
		\centering
		\small{\textbf{\authorThreeName}}\\
		\footnotesize
		\authorThreeAffil
	\end{textblock}
	\begin{textblock}{\authorCellWidth}[0,0](\coloneX,\rowtwoY)
		\centering
		\small{\textbf{\authorFourName}}\\
		\footnotesize
		\authorFourAffil
	\end{textblock}	
	\begin{textblock}{\authorCellWidth}[0,0](\coltwoX,\rowtwoY)
		\centering
		\small{\textbf{\authorFiveName}}\\
		\footnotesize
		\authorFiveAffil
	\end{textblock}	
	\begin{textblock}{\authorCellWidth}[0,0](\colthreeX,\rowtwoY)
		\centering
		\small{\textbf{\authorSixName}}\\
		\footnotesize
		\authorSixAffil
	\end{textblock}
	\begin{textblock}{\authorCellWidth}[0,0](\coloneX,\rowthreeY)
		\centering
		\small{\textbf{\authorSevenName}}\\
		\footnotesize
		\authorSevenAffil
	\end{textblock}	
	\begin{textblock}{\authorCellWidth}[0,0](\coltwoX,\rowthreeY)
		\centering
		\small{\textbf{\authorEightName}}\\
		\footnotesize
		\authorEightAffil
	\end{textblock}	
	\begin{textblock}{\authorCellWidth}[0,0](\colthreeX,\rowthreeY)
		\centering
		\small{\textbf{\authorNineName}}\\
		\footnotesize
		\authorNineAffil
	\end{textblock}
	\begin{textblock}{\authorCellWidth}[0,0](\colDoubleLeftX,\rowfourY)
		\centering
		\small{\textbf{\authorTenName}}\\
		\footnotesize
		\authorTenAffil
	\end{textblock}
	\begin{textblock}{\authorCellWidth}[0,0](\colDoubleRightX,\rowfourY)
		\centering
		\small{\textbf{\authorElevenName}}\\
		\footnotesize
		\authorElevenAffil
	\end{textblock}
}

\defcase{13}{
	\begin{textblock}{\authorCellWidth}[0,0](\coloneX,\rowoneY)
		\centering
		\small{\textbf{\authorOneName}}\\
		\footnotesize
		\authorOneAffil
	\end{textblock}	
	\begin{textblock}{\authorCellWidth}[0,0](\coltwoX,\rowoneY)
		\centering
		\small{\textbf{\authorTwoName}}\\
		\footnotesize
		\authorTwoAffil
	\end{textblock}	
	\begin{textblock}{\authorCellWidth}[0,0](\colthreeX,\rowoneY)
		\centering
		\small{\textbf{\authorThreeName}}\\
		\footnotesize
		\authorThreeAffil
	\end{textblock}
	\begin{textblock}{\authorCellWidth}[0,0](\coloneX,\rowtwoY)
		\centering
		\small{\textbf{\authorFourName}}\\
		\footnotesize
		\authorFourAffil
	\end{textblock}	
	\begin{textblock}{\authorCellWidth}[0,0](\coltwoX,\rowtwoY)
		\centering
		\small{\textbf{\authorFiveName}}\\
		\footnotesize
		\authorFiveAffil
	\end{textblock}	
	\begin{textblock}{\authorCellWidth}[0,0](\colthreeX,\rowtwoY)
		\centering
		\small{\textbf{\authorSixName}}\\
		\footnotesize
		\authorSixAffil
	\end{textblock}
	\begin{textblock}{\authorCellWidth}[0,0](\coloneX,\rowthreeY)
		\centering
		\small{\textbf{\authorSevenName}}\\
		\footnotesize
		\authorSevenAffil
	\end{textblock}	
	\begin{textblock}{\authorCellWidth}[0,0](\coltwoX,\rowthreeY)
		\centering
		\small{\textbf{\authorEightName}}\\
		\footnotesize
		\authorEightAffil
	\end{textblock}	
	\begin{textblock}{\authorCellWidth}[0,0](\colthreeX,\rowthreeY)
		\centering
		\small{\textbf{\authorNineName}}\\
		\footnotesize
		\authorNineAffil
	\end{textblock}
	\begin{textblock}{\authorCellWidth}[0,0](\coloneX,\rowfourY)
		\centering
		\small{\textbf{\authorTenName}}\\
		\footnotesize
		\authorTenAffil
	\end{textblock}	
	\begin{textblock}{\authorCellWidth}[0,0](\coltwoX,\rowfourY)
		\centering
		\small{\textbf{\authorElevenName}}\\
		\footnotesize
		\authorElevenAffil
	\end{textblock}	
	\begin{textblock}{\authorCellWidth}[0,0](\colthreeX,\rowfourY)
		\centering
		\small{\textbf{\authorTwelveName}}\\
		\footnotesize
		\authorTwelveAffil
	\end{textblock}		
	\begin{textblock}{\authorCellWidth}[0,0](\coltwoX,\rowfiveY)
		\centering
		\small{\textbf{\authorThirteenName}}\\
		\footnotesize
		\authorThirteenAffil
	\end{textblock}	
}

\defcase{14}{
	\begin{textblock}{\authorCellWidth}[0,0](\coloneX,\rowoneY)
		\centering
		\small{\textbf{\authorOneName}}\\
		\footnotesize
		\authorOneAffil
	\end{textblock}	
	\begin{textblock}{\authorCellWidth}[0,0](\coltwoX,\rowoneY)
		\centering
		\small{\textbf{\authorTwoName}}\\
		\footnotesize
		\authorTwoAffil
	\end{textblock}	
	\begin{textblock}{\authorCellWidth}[0,0](\colthreeX,\rowoneY)
		\centering
		\small{\textbf{\authorThreeName}}\\
		\footnotesize
		\authorThreeAffil
	\end{textblock}
	\begin{textblock}{\authorCellWidth}[0,0](\coloneX,\rowtwoY)
		\centering
		\small{\textbf{\authorFourName}}\\
		\footnotesize
		\authorFourAffil
	\end{textblock}	
	\begin{textblock}{\authorCellWidth}[0,0](\coltwoX,\rowtwoY)
		\centering
		\small{\textbf{\authorFiveName}}\\
		\footnotesize
		\authorFiveAffil
	\end{textblock}	
	\begin{textblock}{\authorCellWidth}[0,0](\colthreeX,\rowtwoY)
		\centering
		\small{\textbf{\authorSixName}}\\
		\footnotesize
		\authorSixAffil
	\end{textblock}
	\begin{textblock}{\authorCellWidth}[0,0](\coloneX,\rowthreeY)
		\centering
		\small{\textbf{\authorSevenName}}\\
		\footnotesize
		\authorSevenAffil
	\end{textblock}	
	\begin{textblock}{\authorCellWidth}[0,0](\coltwoX,\rowthreeY)
		\centering
		\small{\textbf{\authorEightName}}\\
		\footnotesize
		\authorEightAffil
	\end{textblock}	
	\begin{textblock}{\authorCellWidth}[0,0](\colthreeX,\rowthreeY)
		\centering
		\small{\textbf{\authorNineName}}\\
		\footnotesize
		\authorNineAffil
	\end{textblock}
	\begin{textblock}{\authorCellWidth}[0,0](\coloneX,\rowfourY)
		\centering
		\small{\textbf{\authorTenName}}\\
		\footnotesize
		\authorTenAffil
	\end{textblock}	
	\begin{textblock}{\authorCellWidth}[0,0](\coltwoX,\rowfourY)
		\centering
		\small{\textbf{\authorElevenName}}\\
		\footnotesize
		\authorElevenAffil
	\end{textblock}	
	\begin{textblock}{\authorCellWidth}[0,0](\colthreeX,\rowfourY)
		\centering
		\small{\textbf{\authorTwelveName}}\\
		\footnotesize
		\authorTwelveAffil
	\end{textblock}		
	\begin{textblock}{\authorCellWidth}[0,0](\coloneX,\rowfiveY)
		\centering
		\small{\textbf{\authorThirteenName}}\\
		\footnotesize
		\authorThirteenAffil
	\end{textblock}	
	\begin{textblock}{\authorCellWidth}[0,0](\colthreeX,\rowfiveY)
		\centering
		\small{\textbf{\authorFourteenName}}\\
		\footnotesize
		\authorFourteenAffil
	\end{textblock}	
}

\defcase{15}{
	\begin{textblock}{\authorCellWidth}[0,0](\coloneX,\rowoneY)
		\centering
		\small{\textbf{\authorOneName}}\\
		\footnotesize
		\authorOneAffil
	\end{textblock}	
	\begin{textblock}{\authorCellWidth}[0,0](\coltwoX,\rowoneY)
		\centering
		\small{\textbf{\authorTwoName}}\\
		\footnotesize
		\authorTwoAffil
	\end{textblock}	
	\begin{textblock}{\authorCellWidth}[0,0](\colthreeX,\rowoneY)
		\centering
		\small{\textbf{\authorThreeName}}\\
		\footnotesize
		\authorThreeAffil
	\end{textblock}
	\begin{textblock}{\authorCellWidth}[0,0](\coloneX,\rowtwoY)
		\centering
		\small{\textbf{\authorFourName}}\\
		\footnotesize
		\authorFourAffil
	\end{textblock}	
	\begin{textblock}{\authorCellWidth}[0,0](\coltwoX,\rowtwoY)
		\centering
		\small{\textbf{\authorFiveName}}\\
		\footnotesize
		\authorFiveAffil
	\end{textblock}	
	\begin{textblock}{\authorCellWidth}[0,0](\colthreeX,\rowtwoY)
		\centering
		\small{\textbf{\authorSixName}}\\
		\footnotesize
		\authorSixAffil
	\end{textblock}
	\begin{textblock}{\authorCellWidth}[0,0](\coloneX,\rowthreeY)
		\centering
		\small{\textbf{\authorSevenName}}\\
		\footnotesize
		\authorSevenAffil
	\end{textblock}	
	\begin{textblock}{\authorCellWidth}[0,0](\coltwoX,\rowthreeY)
		\centering
		\small{\textbf{\authorEightName}}\\
		\footnotesize
		\authorEightAffil
	\end{textblock}	
	\begin{textblock}{\authorCellWidth}[0,0](\colthreeX,\rowthreeY)
		\centering
		\small{\textbf{\authorNineName}}\\
		\footnotesize
		\authorNineAffil
	\end{textblock}
	\begin{textblock}{\authorCellWidth}[0,0](\coloneX,\rowfourY)
		\centering
		\small{\textbf{\authorTenName}}\\
		\footnotesize
		\authorTenAffil
	\end{textblock}	
	\begin{textblock}{\authorCellWidth}[0,0](\coltwoX,\rowfourY)
		\centering
		\small{\textbf{\authorElevenName}}\\
		\footnotesize
		\authorElevenAffil
	\end{textblock}	
	\begin{textblock}{\authorCellWidth}[0,0](\colthreeX,\rowfourY)
		\centering
		\small{\textbf{\authorTwelveName}}\\
		\footnotesize
		\authorTwelveAffil
	\end{textblock}		
	\begin{textblock}{\authorCellWidth}[0,0](\coloneX,\rowfiveY)
		\centering
		\small{\textbf{\authorThirteenName}}\\
		\footnotesize
		\authorThirteenAffil
	\end{textblock}	
	\begin{textblock}{\authorCellWidth}[0,0](\coltwoX,\rowfiveY)
		\centering
		\small{\textbf{\authorFourteenName}}\\
		\footnotesize
		\authorFourteenAffil
	\end{textblock}	
	\begin{textblock}{\authorCellWidth}[0,0](\colthreeX,\rowfiveY)
		\centering
		\small{\textbf{\authorFifteenName}}\\
		\footnotesize
		\authorFifteenAffil		
	\end{textblock}
}

\newcommand\authorOneName{Andreas Brunnert\textsuperscript{1}, Andr\'{e} van Hoorn\textsuperscript{2}, Felix Willnecker\textsuperscript{1}, \\%
Alexandru Danciu\textsuperscript{1}, Wilhelm Hasselbring\textsuperscript{3},\\ %
Christoph Heger\textsuperscript{4}, Nikolas Herbst\textsuperscript{5}, Pooyan Jamshidi\textsuperscript{6},\\ %
Reiner Jung\textsuperscript{3}, Joakim von Kistowski\textsuperscript{5}, %
Anne Koziolek\textsuperscript{7},\\ %
Johannes Kro{\ss}\textsuperscript{1}, Simon Spinner\textsuperscript{5}, Christian V{\"o}gele\textsuperscript{1},\\ %
J{\"u}rgen Walter\textsuperscript{3}, Alexander Wert\textsuperscript{4}\\\quad}
\newcommand\authorOneAffil{
	\textsuperscript{1} fortiss GmbH, M{\"u}nchen, Germany\\
	\textsuperscript{2} University of Stuttgart, Stuttgart, Germany\\
	\textsuperscript{3} Kiel University, Kiel, Germany\\
	\textsuperscript{4} NovaTec Consulting GmbH, Leinfelden-Echterdingen, Germany\\
	\textsuperscript{5} University of W{\"u}rzburg, W{\"u}rzburg, Germany\\
	\textsuperscript{6} Imperial College London, London, United Kingdom\\
	\textsuperscript{7} Karlsruhe Institute of Technology, Karlsruhe, Germany\\\quad\\
}

\begin{document}
 
\selectlanguage{english} 
\frontmatter

%%%%%%%%%% Glossary
% \newglossaryentry{label}
% {
%   name={name},
%   description={Some explanation\ldots!!!}
% }

%%%%%%%%%%%%%%%%%%%%%%%%%%%%%%%%%%%%%%%%%%%%%%%%%%%%%%%%%%%%%%%%%%%%%%%%%%%%%%%

% \newacronym{label}{Acronym}{Full text}

\newacronym{API}{API}{Application Programming Interface}
\newacronym{DML}{DML}{Descartes modeling language}
\newacronym{LQN}{LQN}{Layered queuing network}
\newacronym{MARTE)}{MARTE)}{Modeling and analysis of real-time and embedded systems}
\newacronym{PCM}{PCM}{Palladio component model}
\newacronym{QN}{QN}{Queuing network}
\newacronym{QPN}{QPN}{Queuing Petri network}
\newacronym{SDL}{SDL}{Specification and description language}
\newacronym{SPT}{SPT}{Schedulability, performance, and time}
\newacronym{SUT}{SUT}{System Under Test}
\newacronym{UML}{UML}{Unified Modeling Language}

%%%%%%%%%% To see the titlepage uncomment the next line
%% titlepage.tex
%%
\thispagestyle{empty}
% coordinates for the bg shape on the titlepage
\newcommand{\changefont}[3]{\fontfamily{#1} \fontseries{#2} \fontshape{#3} \selectfont}
\newcommand{\diameter}{20} 
\newcommand{\xone}{-25}
\newcommand{\xtwo}{165}
\newcommand{\yone}{20}
\newcommand{\ytwo}{-253}

%%%%%%%%%%%%%%%%%%%%%%%%%
%%%%% Authors array %%%%%
%%%%%%%%%%%%%%%%%%%%%%%%%
\newcommand{\rowoneY}{5.5}
\newcommand{\rowtwoY}{7.0}
\newcommand{\rowthreeY}{8.5}
\newcommand{\rowfourY}{10.1}
\newcommand{\rowfiveY}{11.7}
\newcommand{\rowsixY}{13.3}

% three authors in line
\newcommand{\coloneX}{2.5}
\newcommand{\coltwoX}{7.45}
\newcommand{\colthreeX}{12.4}

% two authors in line
\newcommand{\colDoubleLeftX}{5}
\newcommand{\colDoubleRightX}{10}

% single author in line
\newcommand{\colsinglecentralX}{5.9}

% width of a box for a single author cell
\newcommand{\authorCellWidth}{4.9}

\begin{titlepage}
% Frame shape
\begin{tikzpicture}[overlay]
\draw[color=gray]  
 (\xone mm, \yone mm) -- (\xtwo mm, \yone mm) arc (90:0:\diameter pt) 
  -- (\xtwo mm + \diameter pt , \ytwo mm) -- (\xone mm + \diameter pt , \ytwo mm) 
 arc (270:180:\diameter pt) -- (\xone mm, \yone mm);
\end{tikzpicture}

\changefont{phv}{m}{n}	% helvetica	

%%%%%%%%%%%%%%%%%%%%%%%%%%%%%%% Title
\begin{textblock}{14}[0,0](3,2.3)
	\centering
	\large{\TRnumber}\\
	\vspace*{1cm}
	\huge{\mytitle}\\
	\vspace*{0.5cm}
	\Large{\WGname}
\end{textblock}
%%%%%%%%%%%%%%%%%%%%%%%%%%%%%%% Red bar
\begin{textblock}{15.5}[0,0](2,5.2)
	\begin{tikzpicture}
		\fill[red!80!brown] (0,0cm) rectangle (19.5cm,0.1cm);
	\end{tikzpicture}
\end{textblock}

%%%%%%%%%%%%%%%%%%%%%%%%%%%%%%%%%%%%%%%%%%%%%%%%%%%%%%%%%%%%%%
%%%%%%%%%%%%%%%%%%%%%%%%%%%%%%% Authors %%%%%%%%%%%%%%%%%%%%%%
%%%%%%%%%%%%%%%%%%%%%%%%%%%%%%%%%%%%%%%%%%%%%%%%%%%%%%%%%%%%%%
\begin{center}
	\putAuthors{\numAuthors}
\end{center}

%%%%%%%%%%%%%%%%%%%%%%%%%%%%%%% Logo
%\begin{textblock}{14}[0,0](3,13)
\begin{textblock}{14}[0,0](3,8.5)
	\vspace{1cm}
	\hfill
	\begin{center}
	\includegraphics[width=0.21\textwidth]{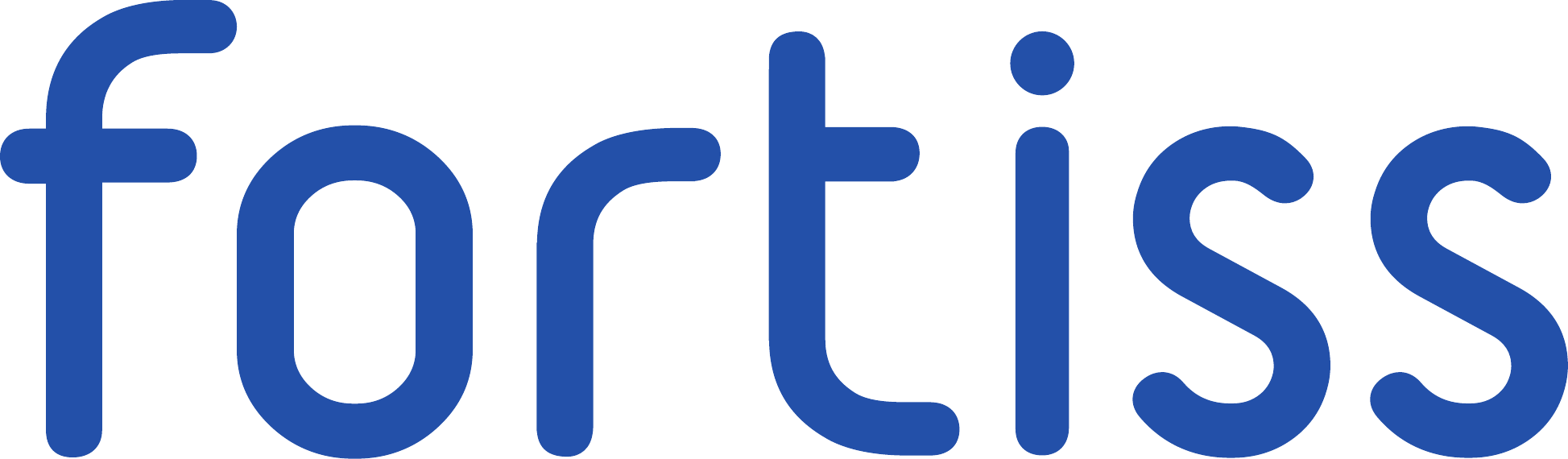}\hfill
	\hspace{1cm}
	\includegraphics[width=0.3\textwidth]{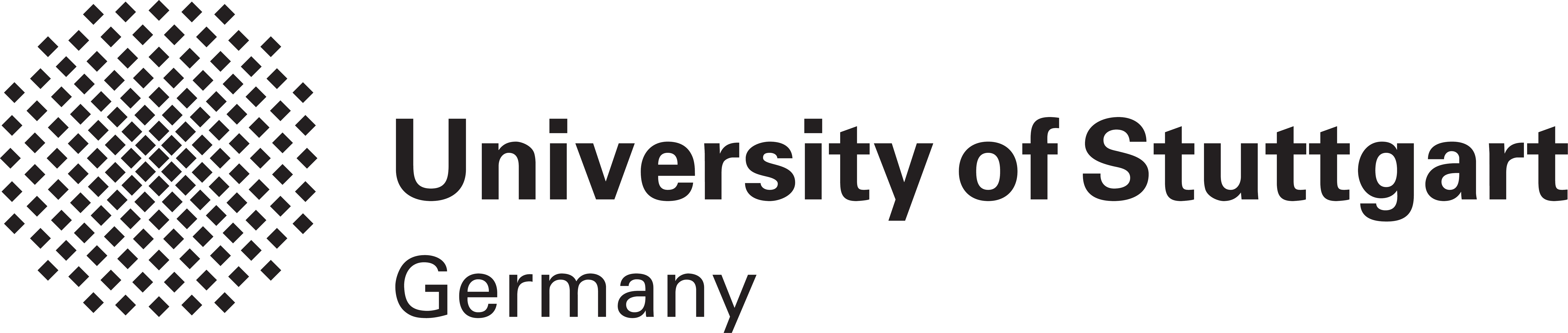}\hfill
	\hspace{1cm}
	\includegraphics[width=0.25\textwidth]{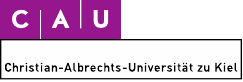}\hfill
	
	\vspace{1,5cm}
	\includegraphics[width=0.25\textwidth]{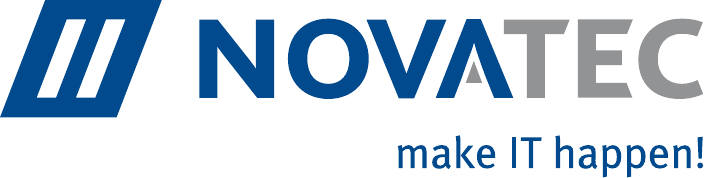}\hfill
	\hspace{1cm}	
	\includegraphics[width=0.22\textwidth]{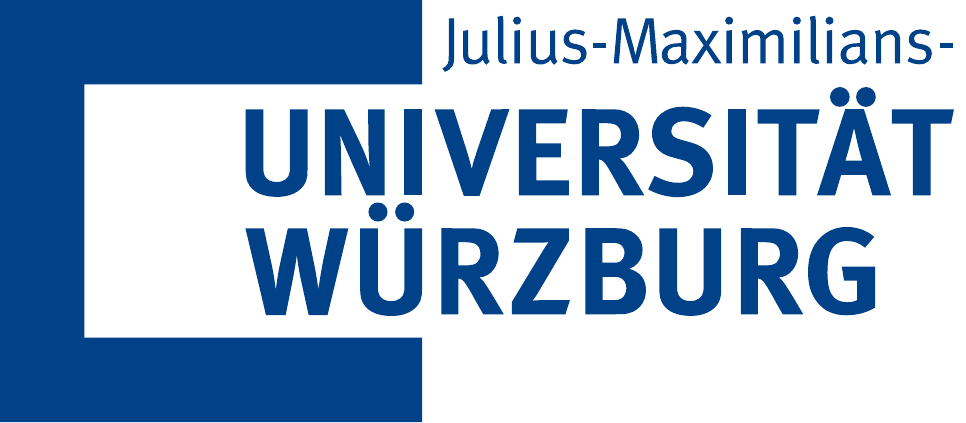}\hfill
	\hspace{1cm}	
	\includegraphics[width=0.25\textwidth]{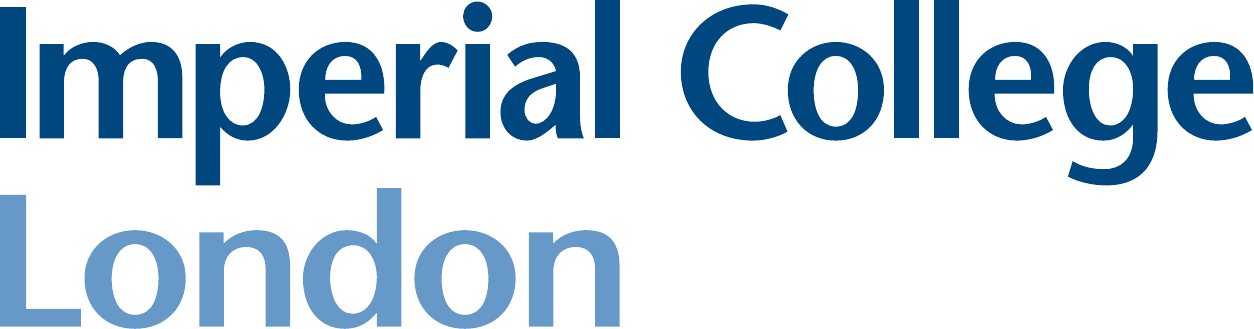}\hfill
	
	\vspace{1,5cm}
	\includegraphics[width=0.2\textwidth]{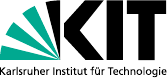}
	\vspace{2cm}
	
	\includegraphics[width=3cm]{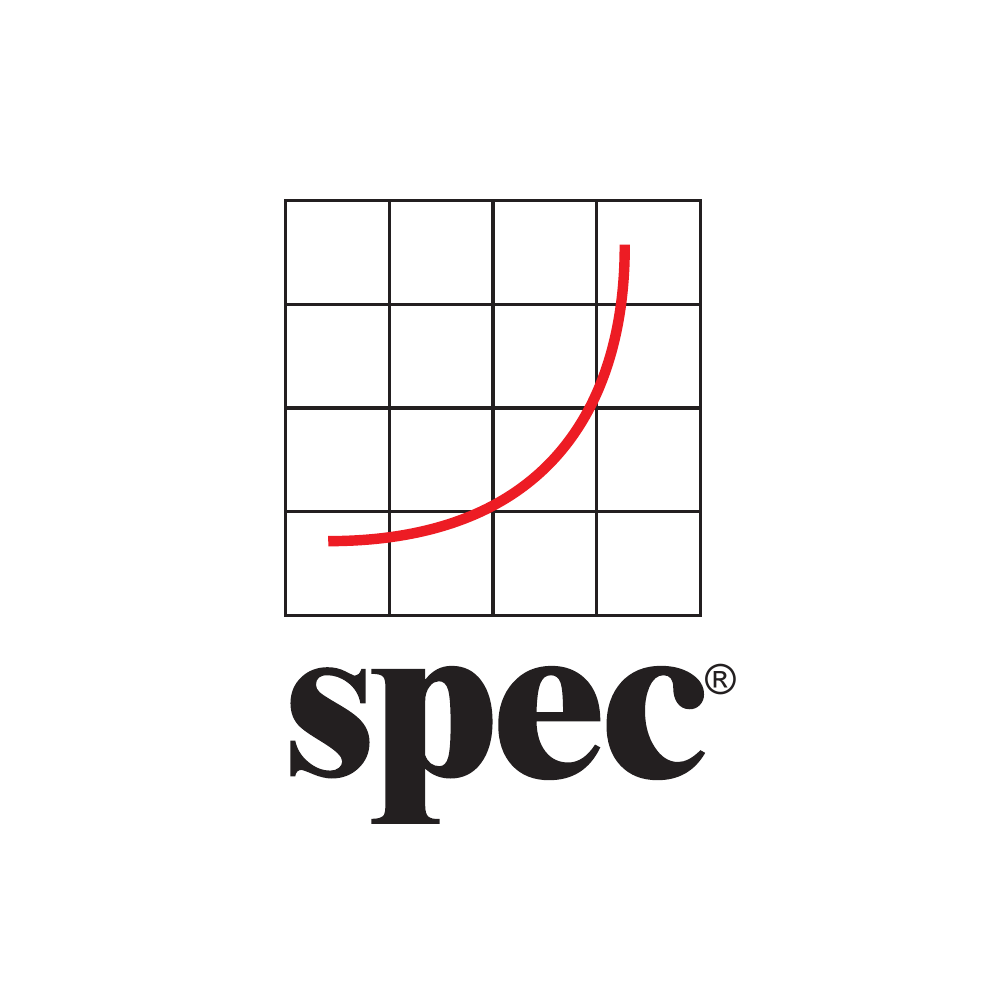} \hfill
	\includegraphics[width=2cm]{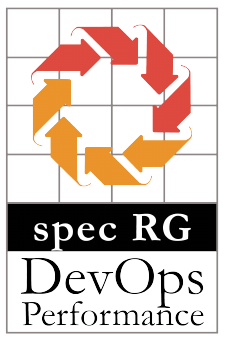} \hfill
	\includegraphics[width=1.9cm]{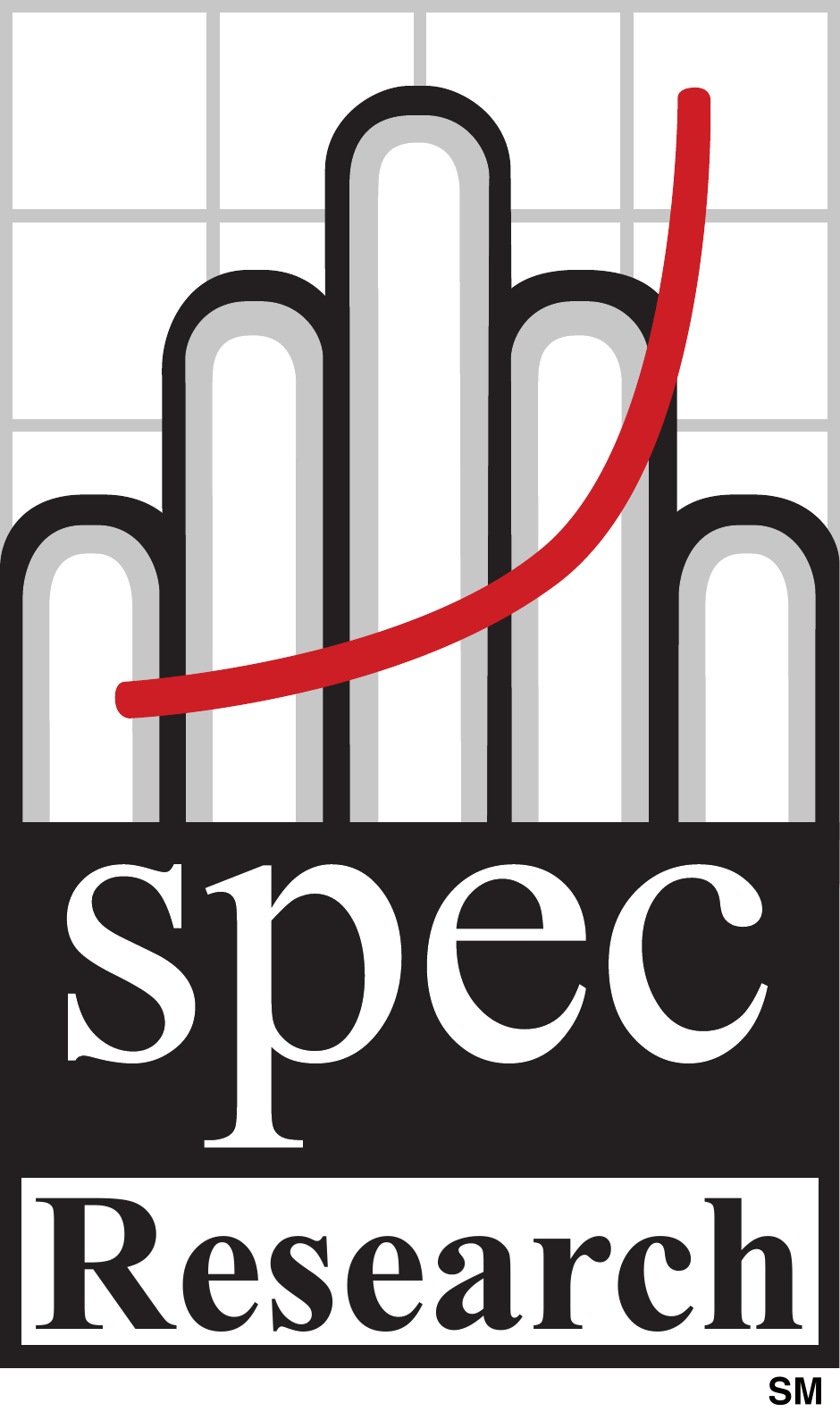} \hspace{1.1cm}\hfill
	\end{center}
	\hfill
\end{textblock}

%%%%%%%%%%%%%%%%%%%%%%%%%%%%%% Acknowledgements
% \begin{textblock}{14}[0,0](3,15)
% 	\noindent\textbf{Acknowledgements}\hfill\vspace{0.5em}\hrule
% 	\vspace{0.5em}\noindent\footnotesize
% 	\Acknowledgements
% \end{textblock}

%%%%%%%%%%%%%%%%%%%%%%%%%%%%%%% Footer
% below the frame
\begin{textblock}{14}[0,0](3,16.75)
	\centering
	\large{\textbf{\TRdate}}
	\hfill
	\large{\textbf{\TRcentralURL}}
	\hfill
	\large{\textbf{\TRrightURL}}
\end{textblock}

\end{titlepage}

%%%%%%%%%% Standard, poor titlepage
%\maketitle
%Alex: One empty page, so when printed out the content is on the left
\newpage 
\thispagestyle{empty}
\mbox{}
%\maketitle

%% -------------------
%% |   Directories   |
%% -------------------
\newpage
\pagenumbering{roman}
\begin{spacing}{1.5}
 \tableofcontents
 \end{spacing}
% \blankpage 

\newpage
\thispagestyle{plain}
\section*{Executive Summary}

DevOps is a trend towards a tighter integration between \ac{Dev} and \ac{Ops} teams. The need for such an integration is driven by the requirement to continuously adapt \acp{EA} to changes in the business environment. As of today, DevOps concepts have been primarily introduced to ensure a constant flow of features and bug fixes into new releases from a functional perspective. In order to integrate a non-functional perspective into these DevOps concepts this report focuses on tools, activities, and processes to ensure one of the most important quality attributes of a software system, namely performance.

Performance describes system properties concerning its timeliness and use of resources. %
Common metrics are response time, throughput, and resource utilization. Performance goals for \acp{EA} are typically defined by setting upper and/or lower bounds for these metrics and specific business transactions. In order to ensure that such performance goals can be met, several activities are required during development and operation of these systems as well as during the transition from \ac{Dev} to \ac{Ops}. Activities during development are typically summarized by the term \ac{SPE}, whereas activities during operations are called \ac{APM}. \ac{SPE} and \ac{APM} were historically tackled independently from each other, but the newly emerging DevOps concepts require and enable a tighter integration between both activity streams. This report presents existing solutions to support this integration as well as open research challenges in this area.

The report starts by defining \acp{EA} and summarizes their characteristics that make performance evaluations for these systems particularly challenging. It continues by describing our understanding of DevOps and explaining the roots of this trend to set the context for the remaining parts of the report. Afterwards, performance management activities that are common in both life cycle phases are explained, until the particularities of \ac{SPE} and \ac{APM} are discussed in separate sections. Finally, the report concludes by outlining activities and challenges to support the rapid iteration between \ac{Dev} and \ac{Ops}.

\acresetall

\vspace{0.5cm}

\noindent \textbf{Keywords} \\
DevOps; Software Performance Engineering; Application Performance Management.

\vspace{0.5cm}

\noindent \textbf{Acknowledgements} \\
\Acknowledgements

\noindent \textbf{} \\
SPEC, the SPEC logo, and the benchmark name SPECjEnterprise are registered trademarks of the Standard Performance Evaluation Corporation (SPEC) and the SPEC Research logo is a service mark of SPEC. Reprint with permission. Copyright \copyright{} 1988--2015 Standard Performance Evaluation Corporation (SPEC). All rights reserved.

%% -----------------
%% |   Main part   |
%% -----------------
\mainmatter
\pagenumbering{arabic}
%\input{schedule.tex}
%%--- section ---%%%

\section{Introduction}
\label{sec:introduction}
\begin{figure}[b!]
	\centering
	\includegraphics[width=0.65\textwidth]{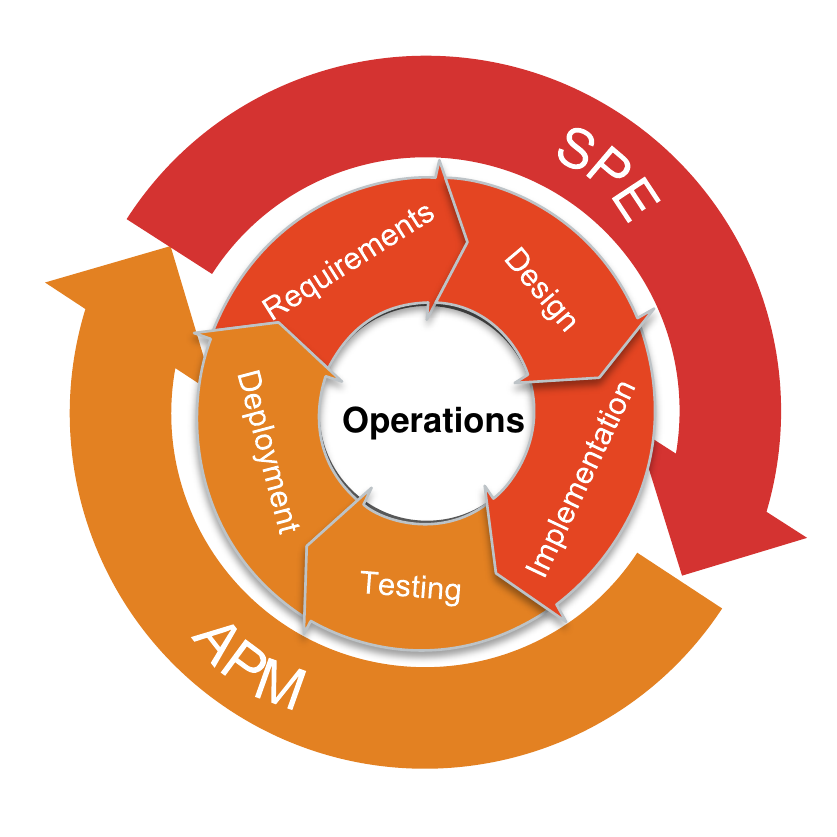}\hfill
	\caption{Performance evaluation in the application life cycle \citep{SPECRGDevOpsWeb2015}}
	\label{fig:ApplicationLifeCycle}
\end{figure}

DevOps has emerged in recent years to enable faster release cycles for complex \ac{IT} services. DevOps is a set of principles and practices for smoothing out the gap between development and operations in order to continuously deploy stable versions of an application system \citep{huettermann2012}. Activities in both of these application life cycle phases often pursue opposing goals. On the one hand, \ac{Ops} teams  want to keep the system stable and favor fewer changes to the system. On the other hand, \ac{Dev} teams try to build and deploy changes to an application system frequently. DevOps therefore aims at a better integration of all activities in software development and operation of an application system life cycle outlined in \autoref{fig:ApplicationLifeCycle}. This liaison reduces dispute and fosters consensus between the conflicting goals of DevOps.

Automation in build, deployment, and monitoring processes are key success factors for a successful implementation of the DevOps concept.
\summary{Automation is DevOps success~fac\-tor}
Technologies and methods used to support the DevOps concept include infrastructure as code, automation through deep modeling of systems, continuous deployment, and continuous integration \citep{PhoenixProject2014}. 

This report focuses on performance-relevant aspects of DevOps concepts. The coordination and execution of all activities necessary to achieve performance goals during system development are condensed as \ac{SPE} \citep{Woodside2007}. Corresponding activities during operations are referred to as \ac{APM} \citep{menasce2004}. Recent approaches integrate these two activities and consider performance management as a comprehensive assignment \citep{brunnert2014}. A holistic performance management supports DevOps by integrating performance-relevant information.

The report summarizes basic concepts of performance management for DevOps and use cases for all phases of an application life cycle. It focuses on technologies and methods in the context of performance management that drive the integration of \ac{Dev} and \ac{Ops}. The report aims on informing and educating DevOps-interested engineers, developers, architects, and managers as well as all readers that are interested in DevOps performance management in general.
In each of the sections representative solutions are outlined. However, we do not claim that all available are covered and are happy to hear about any solution that we have missed (find our contact details on \url{http://research.spec.org/devopswg})---especially, if they address some of the open challenges covered in this report.

After introducing the context of the report in \autoref{sec:context}, the remaining structure is aligned to the different phases of an application life cycle. The underlying functionalities and activities to measure and predict the performance of application systems are explained in \autoref{sec:crossFunctional}. \autoref{sec:designPhase} outlines performance management of DevOps activities in the development phase of an application system. \autoref{sec:integrationPhase} presents performance management  DevOps activities in the operations phase. \autoref{sec:evolutionPhase} describes how performance management can assist and improve the evolution of application systems after the initial roll-out. The report concludes with a summary and highlights challenges for further DevOps integration and performance management.

\section{Context}
\label{sec:context}

This section provides information about the general context of this technical report. \autoref{sec:eaperformance} will highlight the specific characteristics of \aclp{EA} from a performance perspective. Furthermore, in \autoref{sec:devops} we will outline the changes driven by the DevOps movement that make a new view on performance management necessary. 

%!TEX root = ../maindoc.tex
\subsection{Enterprise Application Performance}
\label{sec:eaperformance}

The whole technical report focuses on a specific type of software systems, namely \acp{EA}. %
This term is used to distinguish our perspective from other domains such as embedded systems or work in the field of high performance computing. \acp{EA} support business processes of corporations. This means that they may perform some tasks within a business process automatically, but are used by end-users at some point. Therefore, they often contain some parts that process data automatically and other parts that exhibit a \ac{UI} and require interactions from humans. In case of \acp{EA} these humans can be employees, partners, or customers. 

According to \citet{Grinshpan:2012:SEA:2207820}, performance-relevant characteristics of \acp{EA} include the following. %
\acp{EA} are vital for corporate business functions, especially the performance of such systems is critical for the execution of business tasks. These systems need to be adapted continuously to an ever-changing environment and need customization in order to adjust to the unique operational practices of an organization. Their architecture represents server farms with users distributed geographically in numerous offices. \acp{EA} are accessed using a variety of front-end programs and must be able to handle pacing workload intensities.

Even though we agree with the view on the performance characteristics of \acsp{EA} as outlined by \citet{Grinshpan:2012:SEA:2207820}, we left some of his points out on purpose. A main difference in our viewpoint is that \acsp{EA} may expose \acp{UI} for customers as websites in the Internet such as in e-commerce companies like Amazon. This perspective is a bit different to the perspective of \citet{Grinshpan:2012:SEA:2207820} as he limits the user amount of \acsp{EA} to a controllable number of employees or partners. Internet-facing websites may be used by an unpredictable number of customers. This characteristic poses specific challenges for capacity planning and management activities. 

Performance of \acsp{EA} is described by the metrics response time, throughput, and resource utilization. Therefore, performance goals are typically defined by setting upper and/or lower bounds for these metrics and specific business transactions. In order to ensure that such performance goals can be met, several activities are required during development and operation of these systems as well as during the transition from Dev to Ops. 
%!TEX root = ../maindoc.tex
\subsection{DevOps}
\label{sec:devops}

DevOps indicates an ongoing trend towards a tighter integration between \acl{Dev} (\acs{Dev}) and \acl{Ops} (\acs{Ops}) teams within or across organizations \citep{PhoenixProject2014}. According to \citet{PhoenixProject2014} the term DevOps was initially coined by Debios and Shafer in 2008 and became widespread used after a Flickr presentation in 2009\footnote{http://itrevolution.com/the-convergence-of-devops/}. The goal of the \ac{Dev} and \ac{Ops} integration is to enable \ac{IT} organizations to react more flexibly to changes in the business environment \citep{DevOpsDummies2015}. As outlined in the previous section, \acsp{EA} support or enable business processes. Therefore, any change in the business environment often leads to changing requirements for an \ac{EA}. This constant flow of changes is not well supported by release cycles of months or years. Therefore, a key goal of the DevOps movement is to allow for a more frequent roll-out of new features and bug fixes in a matter of minutes, hours, or days.

This change can be supported by organizational and technical means. From an organizational perspective, the tighter integration of \ac{Dev} and \ac{Ops} teams can be realized by restructuring an organization. This can, for example, be achieved by setting up mixed \ac{Dev} and \ac{Ops} teams for single \acsp{EA} that have end-to-end responsibility for the development and roll-out of an EA. Another example would be to set integrated (agile) processes in place that force \ac{Dev} and \ac{Ops} teams to work closer together. From a technical perspective this integration can be supported by automating as many routine tasks as possible. These routine tasks include things such as compiling the code, deploying new \ac{EA} versions, performing regression tests, and moving an \ac{EA} version from test systems to a production environment. For such purposes, \ac{CI} systems have been introduced and are now extended to \ac{CD} or \ac{CDE} systems \citep{Humble:2010:CDR:1869904}. The differentiation between \ac{CI}, \ac{CD}, and \ac{CDE} is mostly done by the amount of tasks 
these systems automate. Whereas CI systems often only compile and deploy a new \ac{EA} version, \ac{CD} systems also automate the testing tasks until an \ac{EA} version that can be used as a release candidate. \ac{CDE} describes an extension to \ac{CD} that automatically deploys a release candidate to production.

Even though the software engineering community in research and practice has already embraced the changes by introducing approaches for \ac{CI}, \ac{CD}, and \ac{CDE}, a performance perspective for these new approaches is still missing. Specifically, the challenges of the two performance domains \ac{SPE} and \ac{APM} are often considered independently from each other \citep{brunnert2014}. In order to support the technical and organizational changes under the DevOps umbrella driven by the need to realize more frequent release cycles, this conventional thinking of looking at performance activities during \ac{Dev} (\ac{SPE}) and during \ac{Ops} (\ac{APM}) independently from each other needs to be changed. This report outlines existing technologies to support the \ac{SPE} and \ac{APM} integration and outlines open challenges.

\section{Performance Management Activities} % State of the Art and Challenges
\label{sec:crossFunctional}

Even though performance management activities have slightly different challenges during Dev and Ops there are a lot of commonalities in the basic methods used. These common methods are outlined in this section. \autoref{sec:performanceMeasurement} starts with the most fundamental performance management activity which is the measurement-based performance evaluation. As measurement-based performance evaluation methods always have the drawback of requiring a system to measure performance metrics, model-based performance evaluation methods have been developed in order to overcome this requirement. Therefore, \autoref{sec:modelBasedPerformancePrediction} focuses on performance modeling methods. Finally, \autoref{sec:performanceModelExtraction} outlines existing approaches to extract performance models and open challenges for performance model extraction techniques.

\subsection{Measurement-Based Performance Evaluation}
\label{sec:performanceMeasurement}

Measurement-based performance evaluation describes the activity of measuring and analyzing performance characteristics from an executing \acs{EA}. Measurement data can be obtained with event-driven and sampling-based techniques~\citep{lilja2005,menasce2001}. Event-driven techniques collect a measurement whenever a relevant event occurs in the system, e.g., invocation of a certain method. Sampling-based techniques collect a measurement at fixed time intervals, e.g., every second. The tools that collect the measurements are called monitors and are divided into hardware monitors (typically part of hardware devices, e.g., \ac{CPU}, memory, and \ac{HDD}) and software monitors. 

Integrating software monitors into an application is called instrumentation. Instrumentation techniques can be categorized into direct code
modification, indirect code modification using aspect-oriented programming, or
compiler modification, or middleware
interception~\citep{jain2001,lilja2005,kiczales1996aspect,menasce2001}.
The instrumentation is classified as static when the instrumentation is done at
design or compile time, and as dynamic if the instrumentation is done at runtime
without restarting the system. 

The instrumentation and the execution of monitors can alter the behavior of the system at runtime. Software monitors can change
the control flow by executing code that is responsible for creating
measurements. They also compete for shared resources like \ac{CPU}, memory, and storage. The impact of the instrumentation on response times and resource utilization is often called measurement overhead. The degree of measurement overhead depends on the instrumentation granularity (e.g., a single method, all methods of an interface, or all methods of a component), the monitoring strategy (event-driven vs.\ sampling-based), instrumentation strategy (static vs.\ dynamic), and also the types and quality of the employed monitors.

What information is of interest and where the information is to be obtained depends on the performance goals and the life cycle phase of an \acs{EA}. During \ac{Dev}, performance metrics are usually derived using performance, load, or stress tests on a test system, whereas measurements can be directly taken from a production system. The specifics of these activities are outlined in the respective sections later in this report. However, it is important to understand that performance measurements are highly dependent on the system and the workload used to collect them. Therefore, results measured on a one system are not directly applicable for another different system. This is also true for different workloads. Therefore, special care needs to be taken when selecting workloads and test systems for measurement-based performance evaluations during \ac{Dev}.

An overview on available commercial performance monitoring tools is given by
Gartner in its annual published report titled ``Gartner's magic quadrant for
application performance monitoring''~\citep{Gartner2014APMQuad}. The current market
leaders are Dynatrace~\citep{dynatrace}, AppDynamics~\citep{appdynamics},
NewRelic~\citep{newrelic}, and Riverbed Technology~\citep{riverbed}. %
Additionally, free and open source performance monitoring tools exist, %
e.g., Kieker~\citep{KiekerICPE2012}. % RoHoMaSoStGiHa08,vanHoorn2009
%Kieker is an available tool designed for continuous monitoring of systems.

Even though a lot of monitoring tools are available, there are still a lot of challenges to overcome when measuring software performance:

\begin{itemize}
\item The configuration complexity of monitoring tools is often very high and requires a lot of expert knowledge.
\item Monitoring tools lack interoperability in particular with respect to data exchange and accessing raw data. 
\item Selecting an appropriate monitoring tool requires a lot of knowledge of the particular features as some capabilities are completely missing from specific monitoring solutions. 
\item Measurement-based performance evaluation during development requires a representative workload and usage profile (operational profile) to simulate users. In many cases, a replication of the productive system is not available.
\item Setting up a representative test system for measurement-based performance evaluations outside of a production environment is often associated with too much effort and cost. 
\item The accuracy of measurement results is highly platform-dependent, the exact same measurement approach on Linux can exhibit completely different results on Windows for example. 
\item Selecting appropriate time frames to keep historical data during operations is quite challenging. If the time period is too short it might happen that important data is lost too fast, if the period is too long a monitoring solution might run into performance problems itself due to the high amount of data it needs to manage.
\end{itemize}

%!TEX root = ../maindoc.tex
\subsection{Model-Based Performance Evaluation}
\label{sec:modelBasedPerformancePrediction}

\begin{figure}%[ht]
	\centering
	\includegraphics[width=0.725\textwidth]{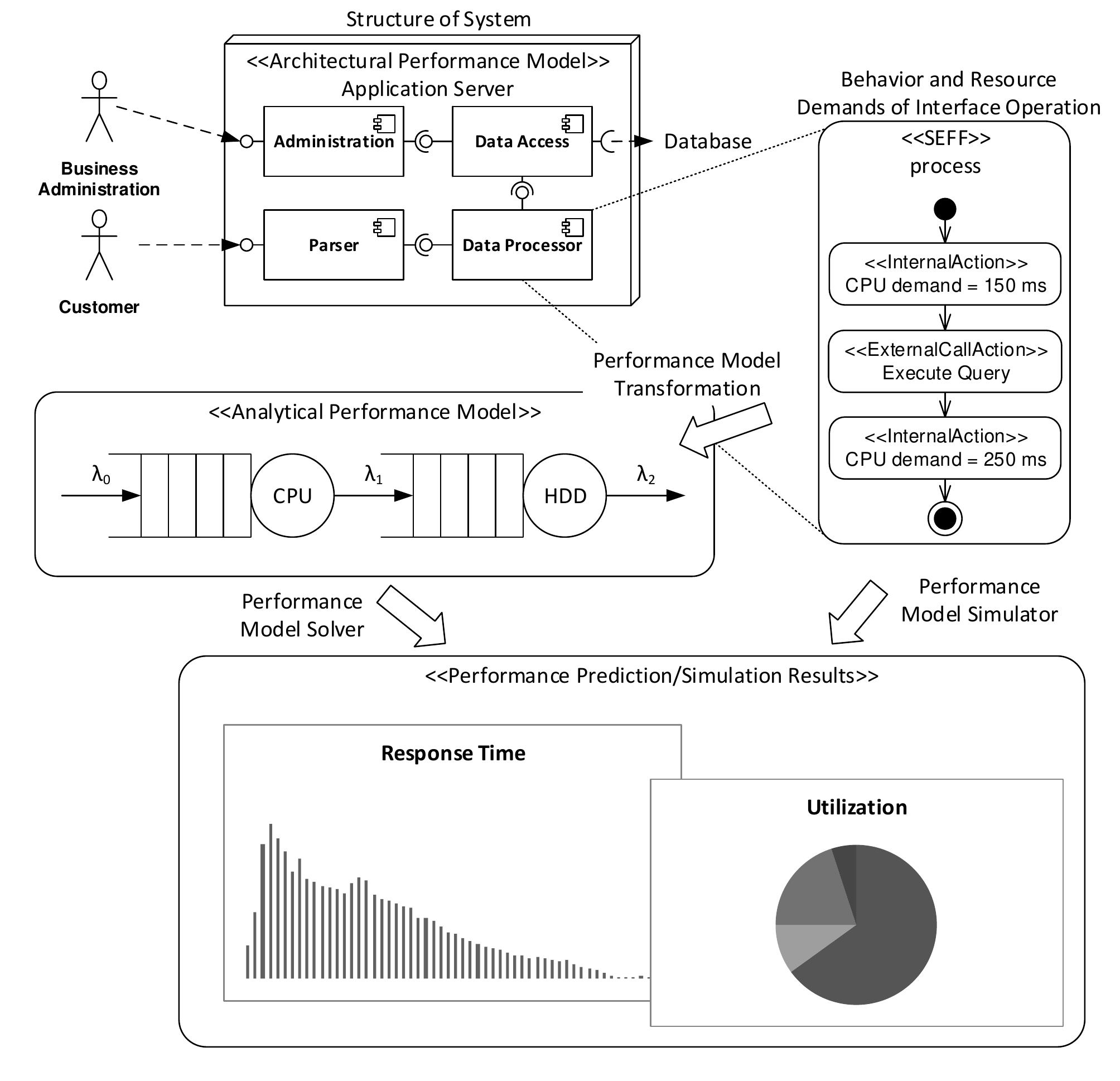}\hfill	
	\caption{Model-based performance evaluation}
	\label{fig:PerformancePrediction}
\end{figure}

Besides the measurement-based approach, performance behavior of a system can be evaluated using  model-based based approaches. %
So-called performance models allow for representing performance-relevant aspects of  software systems and serve as input for analytical solvers or simulation engines. Model-based approaches enable developers to predict performance metrics. This capability can be applied for various use cases within the life cycle of a software system, e.g., for capacity planning or ad-hoc analyses. The procedure is depicted in \autoref{fig:PerformancePrediction}. 

There are two forms of performance models available: analytical models and architecture-level performance models. Common analytical models include Petri nets, \acp{QN}, \acp{QPN}, or \acp{LQN} \citep{balsamo2004,Ardagna2014QoSInCloudModeling}. %They can be used directly for prediction. %
Architecture-level performance models depict key performance-influencing factors of a system's architecture, resources, and usage \citep{Brosig2011}. %
The \acl{UML-SPT} (\acs{UML-SPT}) \citep{OMG2005UMLProfileForSchedulabilityPerformanceAndTimeV1-1}, \acused{UML-SPT} %
the \acl{MARTE} (\acs{MARTE}) \citep{OMG2011UMLProfileForMARTE11}, the \ac{PCM} \citep{becker2009}, \acused{MARTE} %
and the \ac{DML} \citep{Kounev2014} are examples for architecture-level performance models. The latter two models  focus on performance evaluation of component-based software systems 
and allow to evaluate the impact of different influencing factors on software components' performance, which are categorized by \citet{koziolek2010} as follows: 
\begin{itemize}
\item Component implementation: Several components can provide the same interface and functionality, but may differ in their response time or resource usage.
\item Required services: The response time of a service depends on the response time of its required services.
\item Deployment platform: Software components can be deployed on various deployment platforms, which consist of different software and hardware layers.
\item Usage profile: The execution time of a service can depend on the input parameter it was invoked with.
\item Resource contention: The execution time of software components depends on the waiting time it takes to contend required resources.
\end{itemize}

\noindent Architecture-level performance models can be either simulated directly or %
automatically translated into analytical models and, then be processed by %
respective solvers. %
For instance, for \ac{PCM} models different simulation engines and transformations %
to analytical performance models such as \acp{LQN} or stochastic regular expressions %
exist. %
Analytical solvers and simulation engines have in common that they allow for predicting %
performance metrics. 
% Response times are calculated for each modeled function call %
% of all specified components, whereas throughput is recorded in the form of the amount %
% of observations for a certain service or resource during a simulation. Resource utilizations are 
% observed for all specified resources such as CPU or HDD. 

Performance models can be created automatically either based on running applications \citep{Brunnert2013, Brosig2009, Brosig2011} or based on design specifications. Regarding the latter approach, they can be derived from a variety of different design specifications such as the \ac{UML} including sequence, activity, and collaboration diagrams \citep{Petriu2002, Petriu2003, Woodside:2005:PUM:1071021.1071022, brunnert2013integrating}, execution graphs \citep{Petriu2002}, use case maps \citep{Petriu2002}, \ac{SDL} \citep{Kerber2001}, or object-oriented specifications of systems like class, interaction, or state transition diagrams based on object-modeling techniques \citep{Cortellessa2000}.

Performance models and prediction of performance metrics provide the basis to analyze various use cases, especially, to support the DevOps approach. For instance, performance models can be created for new systems during system development which intend to replace legacy systems. Their predicted metrics can then be compared with monitoring data of the existing systems from \ac{IT} operations and, e.g., allow for examining whether the new system is expected to require less resources. Alternatively, performance models of existing systems can be automatically derived from \ac{IT} operations, for example, using the approach by \citet{Brunnert2013} for \ac{Java EE} applications. Subsequently, design alternatives can be evaluated regarding component specifications, software configurations, or system architectures. This enables architects and developers to optimize an existing system for different purposes like efficiency, performance, but also costs and reliability \citep{aleti2013a}. System developers are also able to 
communicate 
performance metrics with \ac{IT} operations and ensure a certain level of system performance across the whole system life cycle. %

Furthermore, model-based performance predictions can be applied to answer sizing questions. System bottlenecks can be found in different places and, for instance, examined already during system development. The ability to vary the workload in a model also allows to evaluate worst-case scenarios such as the impact of an increased number of users on a system in case of promotional actions. In this way, a system's scalability can be examined as well by specifying increased data volumes that have to be handled by components as it may be the case in the future.

Regarding the DevOps approach to combine and integrate activities from software development and \ac{IT} operations, model-based performance evaluation %
is, for instance, useful for %
\begin{inlineenumerate}
\item exchanging and comparing performance metrics during the whole system lifecycle, %
\item optimizing system design and deployment for a given production environment, and %during system development
\item early performance estimation during system development. %
\end{inlineenumerate}

\

Selected challenges for model-based performance prediction include the following: %

\begin{itemize}
\item The representation of main memory as well as garbage collection %
      is not explicitly integrated and considered in performance models, yet. %
\item The selection of appropriate solution techniques %
      requires a lot of expertise. %
\end{itemize}

\subsection{Performance and Workload Model Extraction}
\label{sec:performanceModelExtraction}

Performance models and workload models have to be created before we can deduce performance metrics. 
This section focuses on the extraction of architectural performance models, as they combine the capabilities of architectural models (e.g., UML) and analytical models (e.g., QN). Analytical models explicitly or implicitly assume resource demand of service execution per resource. However, they do not provide a natural linking of resource demands with software elements (e.g., components, operations) like architecture-level performance models useful for DevOPs process automation. 
In traditional long-term design scenarios models may by extracted by hand. However, manual extraction is expensive, error-prone and slow compared to automatic solutions. Especially in contexts where Dev and Ops merge and the models frequently change, automation is of great importance.
The main goal of performance model extraction for DevOps is to define and build an automated extraction process for architectural performance models. Basically, architectural performance models provide a common set of features which have to be extracted. We propose to structure the extraction into the following three extraction disciplines:
%This section outlines existing techniques that and where open challenges remain.

\begin{enumerate}
	\item System structure and behavior,
	\item Resource demand estimation, 		% What about direct measurement?
	\item Workload Characterization. %Usage profile	
\end{enumerate}
These extraction disciplines can be combined to a complete extraction process and are explained in the subsequent sections. Before, we will perform a dissociation of existing model extraction approaches and outline general challenges.
Some predictive models estimate service times without linking resource demands to resources. Approaches targeting their extraction of such black-box models using, for example, genetic optimization techniques (e.g., \citet{Westermann2012} and \citet{Courtois2000}) are not considered in this report. These models serve as interpolation of the measurements. Neither a representation of the system architecture nor its performance-relevant factors and dependencies are extracted.
%\cite{Desnoyders} \todo{describe Desnoyders paper}. 
Approaches to automatically construct analytical performance models, such as QN, have been proposed, e.g., by \citet{Menasce2005,Menasce2007} and \citet{Mos2004}. However, the extracted models are rather limited since they abstract the system at a very high level without taking into account its architecture and configuration. Moreover, for the extraction often imposes restrictive assumptions such as a single workload class or homogeneous servers. Others, like \citet{Kounev2011}, assume the model structure to be fixed and preset (e.g., modeled by hand), and only derive model parameters using runtime monitoring data. Moreover, extraction software is often limited to a certain technology stack (e.g., Oracle WebLogic Server \citep{Brosig2009}). 
We identify the following challenges and goals for future research on model extraction: 	%Resulting 
\begin{itemize}
	\item The assessment of validity and accuracy of extracted models is often based on a trial and error. An improvement would be to equip models with confidence intervals.  % to test if the extracted model predicts accurately.
	\item Model accuracy may expire if they are not updated on changes. Detection mechanisms are required to learn when models get out of date and when to update them.
% The question is about updating frequency. %the model to be updated again?
	\item Current performance modeling formalisms barely ensure the traceability between the running system and model instances. With reference to DevOps, more traceability information should be stored within the models.
	\item The automated inspection of the \acl{SUA} often requires technology-specific solutions. One solution,  to enable less technology-dependent extraction tools, might be self-descriptive resources using standardized interfaces. 
%As it does not provide self description via standardized interfaces.	
%Technology Dependency:   extraction of model aspects hardware, servers,
%	\item Often extraction solutions provide no modular structure.
%Interchangeability/Interoperability of Tools: Common interfaces to exchange components of the extraction process are missing.

	\item The extraction of performance capabilities is based on a combination of software and the (hardware) resources it is deployed at. This combined approach supports prediction accuracy but is less qualified regarding portability of insights to other platforms. One future research direction might be to extract separate models (e.g., separate middleware and application models). 
	\item Automated identification of an appropriate model granularity level. 
	\item Automated identification and extraction of parametric dependencies in call paths and resource demands. % Maybe reference \cite{DissKlausKrogmann}
\end{itemize}

\subsubsection{System Structure and Behavior}
\label{sec:system}

The extraction of structure and behavior describes the configuration of the system.	%
We subdivide the extraction into the extraction of
\begin{inlineenumerate}
\item software components, %
\item resource landscape and deployment, and %
\item inter-component interactions.  
\end{inlineenumerate}
%\paragraph{Software Components}

Software systems that are assembled by predefined components may be represented by the same components in a performance model \cite{Wu2004}. Predefined components (by the  developer) are for example: web services, EJBs in Java EE applications (e.g., in  \cite{Brunnert2013, Brosig2009, Brosig2011}), \texttt{IComponent} extensions in .NET or CORBA components. 
Those extraction techniques depend on predefined components. Existing approaches for software component extraction independent of predefined components target at source code refactoring in a classical development process. Examples for such reverse engineering tools and approaches are for example FOCUS (\cite{Ding2001}), ROMANTIC (\cite{Chardigny2008, Huchard2010}), Archimetrix (\cite{Detten2012}) or 
%ArchiRec\cite{chouambe2008} / 
SoMoX (\cite{Becker2010, Krogmann2010}). These approaches are either clustering-based, pattern-based, or combine both. They all identify components as they should be according to several software metrics. However, this does not necessarily correspond to the actual deployable structures, which is required during operation. Consequently, the identified components may be deployable in multiple parts. The reverse engineering approaches satisfy the 'Dev' but not the 'Ops' part of DevOps. An automated approach that works for DevOps independent of predefined component definitions is still an open issue. If no predefined components are provided, component definition requires manual effort. 
%In general, \todo{Here should be a source} 
For manual extraction the following guidelines can be applied: i) classes that inherit from component interfaces (e.g., IComponent or EJBComponent) represent components, ii) all classes that inherit from a base class belong to the same component, iii) if component A uses component B then A is a composite component including B.
% Christoph Heger: Warum ist es dann eine composite component? Ist doch nur eine Abhängigkeit. Composite Beziehung erschließt sich mir nicht?
Component extraction has the major challenges of technology dependency. Currently no tool that covers a wider range of component technologies is known to the authors. 
%\begin{itemize}
%	\item Current work on extraction is semi-automatically focusing one technological stack. %using research prototypes. It is still an open issue to integrate component identification for different %technologies in one tool \todo{HereShouldBeAReference}. 
%\end{itemize}

%\paragraph{Resource Landscape and Deployment}
Besides software component identification one has to extract resource landscape and deployment. Automated identification of hardware and software resources in a system environment is already available in industry. For instance, \citet{hyperic2014} or \citet{zenoss214} provide such functionalities. Given a list of system elements, system, network and software properties can be extracted automatically. Further,  low-level aspects, like cache topology, can be extracted using open source tools like LIKWID (\cite{Treibig2010}). 
%Further, network infrastructure extraction approaches exist es well. 
%However, all approaches visualize extracted information within the extraction tool. An open issue is to define common interfaces for resource extraction. This would greatly improve the integration into %tool chains and guarantee interoperability.
% \todo{Add: \todo{Hintergrund war, das wir in der Situation darüber gesprochen hatten, in wie fern man die Performance-Charakteristika einer Runtime Umgebung (z.B. Java EE Server) unabhängig von der %Komponentenbeschreibung beschreiben kann.}
%Approaches for representing middleware independently from the application: Chen et al. [CLGL05] derive mathematical models from measurements to create product-specific performance profiles for %the EJB runtime of a Java EE server. These models are intended to be used for performance predictions of EJB components running on different Java EE products. \cite{1556552} %==Liu2005}
%\todo{Clarify the following two sentences}
The deployment, which is the mapping of software to resources, can be extracted using service event logs. These logs usually contain for an executed operation (besides the execution time) identifiers that enable a mapping to the corresponding software component and the machine it was executed at. The extraction happens by the creation of one deployment component per couple of software component and resource identifier found within the event logs. The logging means no additional effort as the logs are also required for resource demand estimation.
% and can be received from tools like DynaTrace, AppDynamics or Kieker.
The extraction of a resource landscape in performance modeling is mostly performed semi-automatically. We ascribe this to mainly technical challenges, e.g.,  
integration of information from different sources with many degrees of freedom (network, CPU count and clock frequency, memory, middleware, operating systems). % 

%\paragraph{Inter-Component Interaction} % / Structural Information
The extraction of interactions between components differs for design time and runtime. At design time, models can be created using designer expertise and design documents (e.g., as performed in \cite{Smith2002, Menasce2000, Petriu2002, Cortellessa2000}). Commencing at a runnable state, monitoring logs can be generated. Automated extraction of structural information based on monitoring logs has the advantage that it tracks the behavior of the actual product as it is evolved. An \emph{effective architecture} can be extracted which means that only executed system elements are extracted (\cite{Israr2007}). Further, runtime monitoring data enables to extract branching probabilities for different call paths (\cite{Brosig2014}). 
Selected approaches for control flow extraction are by \cite{Hrischuk1999, Briand2006} and \cite{Israr2007}. \cite{Hrischuk1999} and \cite{Briand2006} use monitoring information based on probes which are injected into the beginning of each response and propagated through the system. 
%A probe is a header information 
The approach of \cite{Israr2007} requires no probe information but is unable to model synchronization delays in parallel sub-paths which join together. In contemporary monitoring tools like DynaTrace, AppDynamics or Kieker the probe-based approach became standard.

%Besides: 
%An approach to ensure synchronization between model and reality is DiscoTect \cite{Yan2004, Schmerl2006}. It compares monitoring data with system specification to detect inconsistencies.

\

We identify the following major challenges for structure and behavior extraction:
\begin{itemize}
	\item Component extraction customization effort for case studies. Portability of technology dependent component extraction approaches is low. Current technology-independent component extraction approaches are considered not to be capable for a fully automated performance model extraction. 
	\item Monitoring customization effort for case studies. A complete extraction story requires a lot of tools with different interfaces to be connected. Especially, the portability and combination of multiple resource extraction approaches is a complex task.
\end{itemize}

\subsubsection{Resource Demand Estimation}
\label{sec:resourceDemandEstimation}

In architecture-level performance models, resource demands are a key parameter to enable their quantitative analysis. A resource demand describes the amount of a hardware resource needed to process one unit of work (e.g., user request, system operations, or internal actions). The granularity of resource demands depends on the abstraction level of the control flow in a performance model. Resource demands may depend on the value of input parameters. This dependency can be either captured by specifying the stochastic distributions of resource demands or by explicitly modeling parametric dependencies. 

The estimation of resource demands is challenging as it requires a deep integration between application performance monitoring solutions and operating system resource usage monitors in order to obtain resource demand values. Operating system monitors often only provide aggregate resource usage statistics on a per-process level. However, many applications (e.g., web and application servers) serve different types of requests with one or more processes.

Profiling tools \citep{Graham1982,Hall1992} are typically used during development to track down performance issues as well as to provide information on call paths and execution times of individual functions. These profiling tools rely on either fine-grained code instrumentation or statistical sampling. However, these tools typically incur high measurement overheads, severely limiting their usage during production, and leading to inaccurate or biased results. In order to avoid distorted measurements due to overheads, \citet{kuperberg2008a,kuperberg2009c} propose a two-step approach. In the first step, dynamic program analysis is used to determine the number and types of bytecode instructions executed by a function. In a second step, the individual bytecode instructions are benchmarked to determine their computational overhead. However, this approach is not applicable during operations and fails to capture interactions between individual bytecode instructions. \ac{APM} tools, such as \citet{dynatrace} or \citet{appdynamics}, enable 
fine-grained monitoring of the control flow of an application, including timings of individual operations. These tools are optimized to be also applicable to production systems.

Modern operating systems provide facilities to track the consumed \ac{CPU} time of individual threads. This information is, for example, also exposed by the Java runtime environment. This information can be exploited to measure the \ac{CPU} resource consumption of processing individual requests as demonstrated for Java by \citet{Brunnert2013} and at the operating system level by \citet{Barham2004}. This requires application instrumentation to track which threads are involved in the processing of a request. This can be difficult in heterogeneous environments using different middleware systems, database systems, and application frameworks. The accuracy of such an approach heavily depends on the accuracy of the \ac{CPU} time accounting by the operating system and the extent to which request processing can be captured through instrumentation.

Over the years, a number of approaches to estimate the resource demands using statistical methods have been proposed. These approaches are typically based on a combination of aggregate resource usage statistics (e.g., \ac{CPU} utilization) and coarse-grained application statistics (e.g., end-to-end application response times or throughput). These approaches do not depend on a fine-grained instrumentation of the application and are therefore widely applicable to different types of systems and applications  incurring only insignificant overheads. Different approaches from queuing theory and statistical methods have been proposed, e.g., response time approximation \citep{Brosig2009,Urgaonkar2007}, least-squares regression~\citep{Bard1978,Rolia1995,Pacifici2008}, robust regression techniques~\citep{Casale2007,Casale2008}, cluster-wise regression~\citep{Cremonesi2010}, Kalman Filter~\citep{Zheng2008,Kumar2009,Wang2012}, optimization techniques~\citep{Zhang2002,Liu2006,Menasce2008,Kumar2009a}, Support Vector 
Machines~\citep{Kalbasi2011}, Independent Component Analysis~\citep{Sharma2008}, Maximum Likelihood Estimation~\citep{Kraft2009,wang2013-MASCOTS,Perez2015EstimatingDemands}, and Gibbs Sampling~\citep{sutton2011,perez2013-MASCOTS}. These approaches differ in their required input measurement data, their underlying modeling assumptions, their output metrics, their robustness to anomalies in the input data, and their computational overhead. A detailed analysis and comparison is provided by \citet{SpCaBrKo2015-PEVA-ResDemEstSurvey}. A \ac{LibReDE} offering ready-to-use implementations of several estimation approaches is described in \citet{spinner2014}.

\

We identify the following areas of future research on resource demand estimation: %
\begin{itemize}
\item Current work is mainly focused on \ac{CPU} resources. More work is required to address the specifics of other resource types, such as memory, network, or \ac{I/O} devices. The challenges with these resource types are, among others, that the utilization metric is often not as clearly defined as for \acp{CPU}, and the resource access may be asynchronous.
\item Comparisons between statistical estimation techniques and measurement approaches are missing. This would help to better understand their implications on accuracy and overhead.
\item Most approaches are focused on estimating the mean resource demand. However, in order to obtain reliable performance predictions it is also important to determine the correct distribution of the resource demands.
\item Modern system features (e.g., multi-core \acp{CPU}, dynamic frequency scaling, virtualization) can have a significant impact on the resource demand estimation. 
\item Resource demand estimation techniques often require measurements for all requests during a certain time period in which a resource utilization is measured, whereas resource demand measurements can be applied for a selected set of transactions. 
\end{itemize}

\subsubsection{Workload Characterization and Workload Model Extraction}%\subsubsection{Workload Characterization}
\label{sec:workloadCharacterization}
{Workload characterization} is a performance engineering activity that serves to %
\begin{inlineenumerate}
\item study the way users (including other systems) interact 
             with the \ac{SUA} via the system-provided 
             interfaces and to 
\item create a workload model that provides an abstract representation 
             of the usage profile \citep{jain2001}.  
\end{inlineenumerate}
             
\citet{menasce2001} suggest to decompose the system's 
{global workload} into {workload components} (e.g., distinguishing web-based interactions from client/server transactions), which are further divided into {basic components}. Basic components (e.g., representing business-to-business transaction types 
or services invoked by interactive user interactions via a web-based \ac{UI}) are assigned {workload intensity} (e.g., arrival rates, number of user sessions over time, and think times) and {service demand} (e.g., average number of documents retrieved per service request) parameters. 

The remainder of this section focuses on the extraction of workload characteristics related to navigational profiles (Section~\ref{sec:UsageModel}) and workload intensity (Section~\ref{sec:loadIntensityProfiles}).  

\paragraph{Navigational Profiles}\label{sec:UsageModel}

For certain kinds of systems, the assumption of workload being an arrival of independent requests is inappropriate. A common type of enterprise applications are {session-based systems}. In these systems, users interact with the system in a sequence of inter-related requests, each being submitted after an elapsed {think time}. The notion of a {navigational profile} is used to refer to the session-internal behavior of users. 
The navigational profile captures the possible ways or states of a workload unit (single user/customer) through the system. Note that we do not limit the scope of the targeted systems to web-based software systems but to multi-user enterprise application systems in general. The same holds for the notion of a {request}, which is not limited to web-based software systems. One goal of the session-based notion is to group types of users with a similar behavior.                           

 \subparagraph*{Metrics and Characteristics.} 
For session-based systems, workloads characteristics can be divided into {intra-session} and {inter-session} characteristics. Intra-session characteristics include think times between requests and the session length, e.g., in terms of the time elapsed between the first and the last request within a session and the number of requests within a session. Inter-session characteristics include the number of sessions per user and the number of active sessions over time as a workload intensity metric. Moreover, request-based workload characteristics apply, e.g., the distribution of invoked request types observed from the server perspective. 

 \subparagraph*{Specification and Execution of Session-Based Workloads.} 

Two different approaches exist to specify session-based workloads, namely based on %
\begin{inlineenumerate}
\item scripts and %
\item on performance models. % 
\end{inlineenumerate}

Script-based  specification is supported by essentially every load testing tool. The workflow of a single user (class) is defined in a programming-language style---sometimes even using programming languages such as Java or C++ (e.g., HP Loadrunner). These scripts, representing a single user, are then executed by a defined number of concurrent load generator threads. Even though the scripting languages provide basic support for probabilistic paths, the scripts are usually very deterministic. 

\begin{figure}
\centering
 	\includegraphics[width=0.6\textwidth]{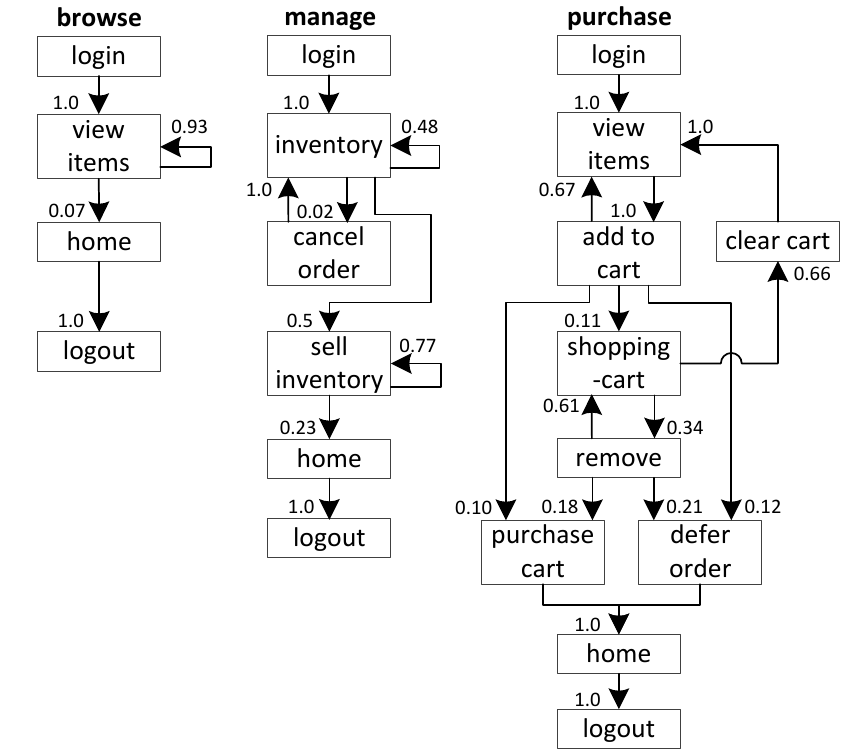} %
 \caption{SPECjEnterprise2010 transaction types (\textit{browse}, \textit{manage}, and \textit{purchase}) in a \acs{CBMG} (Markov chain) representation \citep{vanHoornVoegeleSchulzHasselbringKrcmar2014AutomaticExtractionOfProbabilisticWorkloadSpecificationsForLoadTestingSessionBasedApplicationSystems}. %
 For examples, in \textit{browse} transactions, users start with a \textit{login}, followed by a \textit{view item} request in $100\%$ of the cases; %
 \textit{view items} is followed by \textit{view items} with a probability of $93\%$ and by \textit{home} in $7\%$ of the cases.}
 \label{fig:cs:workloadModel:illustrative}
\end{figure}

As opposed to this, performance models provide an abstract representation of a user session---usually including probabilistic concepts. An often-used formalism for representing navigational profiles in session-based systems are Markov chains 
\citep{MenasceAlmeidaFonsecaMendes1999AMethodologyForWorkloadCharacterizationOfECommerceSites,vanHoornRohrHasselbring2008GeneratingProbabilisticAndIntensityVaryingWorkloadForWebBasedSoftwareSystems,LiTian2003TestingTheSuitabilityOfMarkovChainsAsWebUsageModels}, i.e., probabilistic finite state machines. For example,  \citet{MenasceAlmeidaFonsecaMendes1999AMethodologyForWorkloadCharacterizationOfECommerceSites} introduce a formalism based on Markov chains, called \acp{CBMG}. In a \ac{CBMG}, states represent possible interactions with the \ac{SUA}. Transitions between states have associated transition probabilities and average think times.
For example, \autoref{fig:cs:workloadModel:illustrative} depicts \acp{CBMG} for transactions types of a %
(modified) workload used by the industry-standard benchmark  SPECjEnterprise2010 %
\citep{vanHoornVoegeleSchulzHasselbringKrcmar2014AutomaticExtractionOfProbabilisticWorkloadSpecificationsForLoadTestingSessionBasedApplicationSystems}. %
\acp{CBMG} can be used for workload generation \citep{Menasce2002TPCW-ABenchmarkForECommerce}. As emphasized by \citet{KrishnamurthyRoliaMajumdar2006ASyntheticWorkloadGenerationTechniqueForStressTestingSession-BasedSystems} and \citet{ShamsKrishnamurthyFar2006AModelBasedApproachForTestingThePerformanceOfWebApplications}, limitations apply 
when using \acp{CBMG} for workload generation. For example, 
the simulation of the Markov chain may lead to violations of inter-request dependencies, i.e., to sequences of requests that do not respect the protocol of the \ac{SUA}. Two items, for instance, may be removed from a shopping cart, even though only a single item has been added before. To support inter-request dependencies (and data dependencies), \citet{ShamsKrishnamurthyFar2006AModelBasedApproachForTestingThePerformanceOfWebApplications} propose a workload modeling approach based on (non-deterministic) \acp{EFSM}. An \ac{EFSM} describes valid sequences of requests within a session. As opposed to \acp{CBMG}, transitions are not labeled with probabilities but with predicates and actions based on pre-defined state variables. The actual workload model is the combination of valid sessions obtained by simulating an \ac{EFSM} along with additional workload characteristics like session inter-arrival times, think times, session lengths, and a workload mix modeling the relative frequency of request types. %
Van Hoorn et al. (\citeyear{vanHoornRohrHasselbring2008GeneratingProbabilisticAndIntensityVaryingWorkloadForWebBasedSoftwareSystems,vanHoornVoegeleSchulzHasselbringKrcmar2014AutomaticExtractionOfProbabilisticWorkloadSpecificationsForLoadTestingSessionBasedApplicationSystems}) combine the aforementioned modeling approaches based on \acp{CBMG} and \acp{EFSM}. Other approaches based on analytical models employ variants of Markov chains \citep{Barber2004UserCommunityModelingLanguageUCML11ForPerformanceTestWorkloads}, Probabilistic Timed Automata \citep{Abbors2013Mbpet} and \ac{UML} State Machines \citep{becker2009,OMG2005UMLProfileForSchedulabilityPerformanceAndTimeV1-1,omgMarte}. 
                
 \subparagraph*{Extraction.} 
   
Extractions of navigational profiles are usually based on {request logs} obtained from a \ac{SUA}. For web-based systems, these logs (also referred to as {access logs} or {session logs}) usually include for each request the identifier of the requested service, a session identifier (if available), and timing information (time of request and duration). Data mining techniques, such as clustering, are used to extract the 
aforementioned formal models  \citep{MenasceAlmeidaFonsecaMendes1999AMethodologyForWorkloadCharacterizationOfECommerceSites,vanHoornVoegeleSchulzHasselbringKrcmar2014AutomaticExtractionOfProbabilisticWorkloadSpecificationsForLoadTestingSessionBasedApplicationSystems}. 

\

We identify the following challenges and future directions for navigational profiles as part of workload characterization: 

\begin{itemize}
 \item Most methods so far have focused on extracting and characterizing navigational profiles offline. Promising future work is to perform such extraction and characterization continuously, e.g., to use the gathered information for near-future predictions which do not only take the workload intensity into account. 
 \item Navigational profiles, or workload scripts in general, outdate very fast due to changes as part of the evolving \ac{SUA}. This concerns the expected usage pattern for the application but also protocol-level details in the interaction (service identifiers, parameters, etc.). A future direction could be to utilize navigational profiles already during development. 
\end{itemize}
  
\paragraph{Load Intensity Profiles} \label{sec:loadIntensityProfiles} 

A load intensity profile definition is a crucial element to complete a workload characterization. The observed or estimated arrival process of transactions (on the level of users, sessions or requests/jobs arrivals) needs to be specified. As basis to specify time-dependent arrival rates or inter-arrival times, the extraction of a usage model (see \autoref{sec:UsageModel}) should provide a classification of transaction types that are statistically indistinguishable in terms of their resource demanding characteristics. A {load intensity profile} is an instance of an arrival process. A workload that consists of several types of transactions is then characterized by a set of load intensity profile instances.

Load intensity profiles are directly applicable in the context of any open workload scenario with a theoretically unlimited number of users, but are not limited to those \citep{schroeder2006}. In a closed or partially closed workload scenario, with a limited number of active transactions, the arrival process can be specified within the given upper limits and zero. Any load intensity profile can be transformed into a time series containing arrival rates per sampling interval.
  
\begin{figure}[ht]
	\centering
	\includegraphics[width=0.7\textwidth]{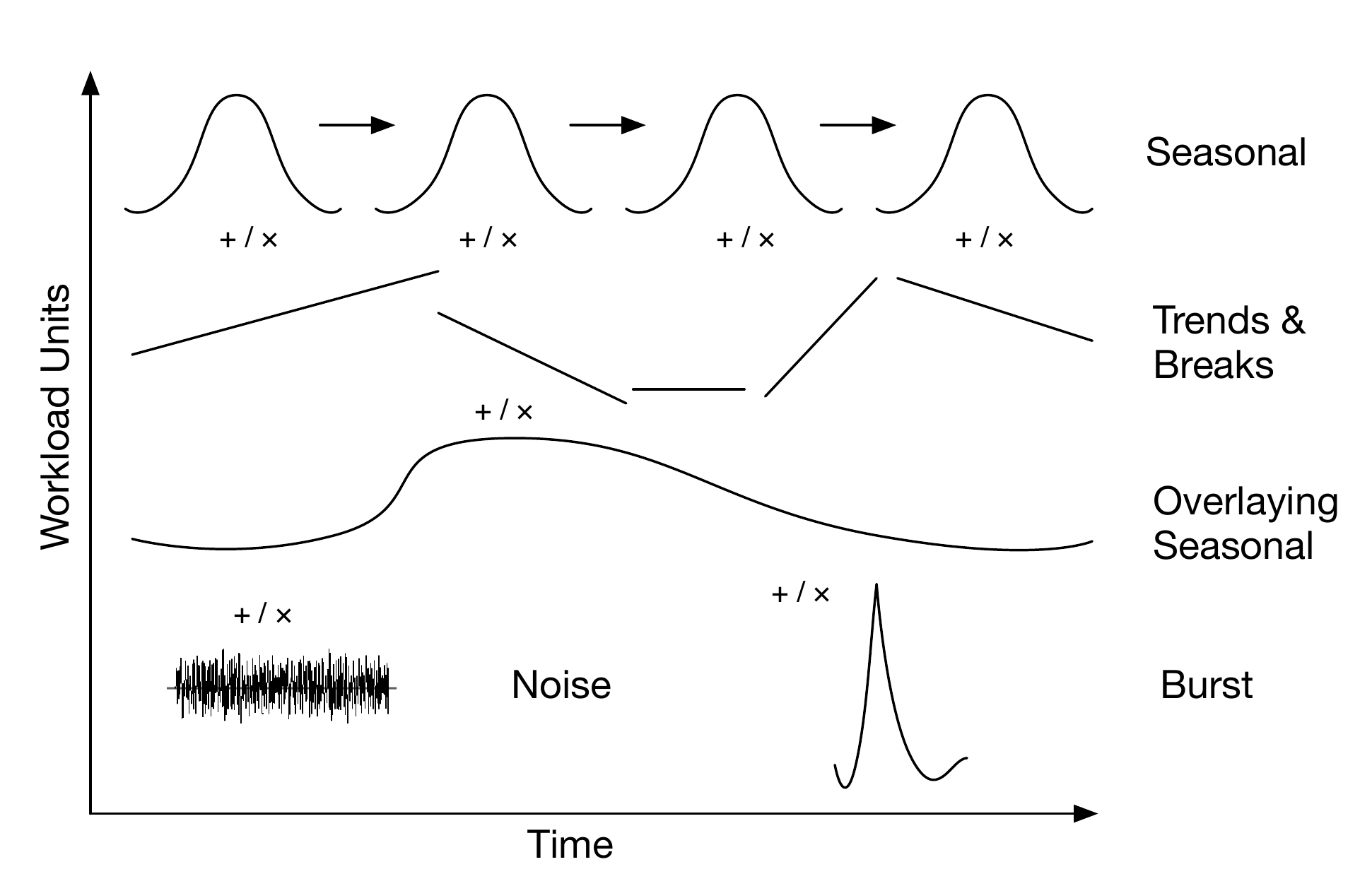}\hfill
	\caption{Elements of load intensity profiles \citep{KiHeKo14}}
	\label{fig:LoadIntensityProfile}
\end{figure}

A requirement for a load profile to appear as realistic (and not synthetic) for a given application domain is a mixture of %
\begin{inlineenumerate}
\item one or more (overlaying) seasonal patterns, %
\item long term trends including trend breaks, %
\item characteristic bursts, and 
\item a certain degree of noise. % 
\end{inlineenumerate}
These components can be combined additive or multiplicative over time as visualized in \autoref{fig:LoadIntensityProfile}.
  
At early development stages, load intensity profiles can be estimated by domain experts by defining synthetic profiles using statistical distributions or mathematical functions. At a higher abstraction level, the \ac{DLIM} allows to descriptively define the seasonal, trends, burst, and noise elements in a wizard-like manner \citep{KiHeKo14}. \ac{DLIM} is supported by a tool-chain named \ac{LIMBO} \citep{LIMBO}. A good starting point for a load intensity profile definition at the development stage is to analyze the load intensity of comparable systems within the same domain.
If traces from comparable systems are available, a load profile model can be extracted in a semi-automated manner as described by \citet{KiHeZoKoHo2015DLIMExtraction}. %

\

We identify the following open challenges in the field of load profile description and their automatic extraction:
\begin{itemize}
    \item Seasonal patterns may overlay (e.g., weekly and daily patterns) and change in their shape over time. The current extraction approaches do not fully support these scenarios.
\end{itemize}

\section{Software Performance Engineering During Development} % (Dev)
\label{sec:designPhase}

This section focuses on how a combination %
of model-based and measurement-based techniques can support %
performance evaluations during software development. %
First, we will focus on the challenges of how to conduct meaningful %
performance analyses in stages where no implementation exists (\autoref{sec:PerformanceModelsRequirementsPhase}). %
During development, timely feedback and guidance on %
performance-relevant properties of implementation decisions %
is extremely valuable to developers, e.g., concerning %
the selection of algorithms or data structures. %
Support for this is provided under the umbrella of %
performance awareness, presented in \autoref{sec:performanceAwareness}. %
Next, in \autoref{sec:performanceAntiPattern}, we present approaches to %
detect performance problems %
automatically based on analyzing design models and runtime %
observations. %  
Finally, \autoref{sec:performanceChangeDetection} focuses on the automatic %
detection of performance %
regressions, including approaches that combine measurements %
and model-based performance prediction. %

\subsection{Design-Time Performance Models}
%\subsection{Performance Models in early Software Development Phases}
\label{sec:PerformanceModelsRequirementsPhase}

In early software development phases like the design phase a lot of architectural, design and technology decisions must be made that can have a significant influence on the performance during operations \citep{koziolek2010}. However, predicting the influence of design decisions on the performance is difficult in early stages. Many questions arise for software developers and architects during these phases. These questions include but are not limited to:

\begin{itemize}
 \item What influence does a specific design decision have on performance? 
 \item How scalable is the designed software architecture? 
 \item Given the performance of reused components/systems, can the performance goals be achieved? 
\end{itemize}

\noindent In the early software development phases, performance models can be used as ``early warning'' instrument to answer these questions \citep{Woodside2007}. The goal of using performance models is to support performance-relevant architecture design decisions. Using performance models in early development phases should also motivate software developers to engage in early performance discussions like answering what-if questions \citep{Thereska2010}. A what-if question describes a specific case for design or architectural decision like: "What happens if we use client-side rendering?" as an alternative to "What happens if we use server-side rendering?". 

In order to create performance models, information about the system's architecture (i.e., scenarios describing system behavior and deployment), workloads (see \autoref{sec:workloadCharacterization}), and resource demands (see \autoref{sec:resourceDemandEstimation}) is needed. However, in early phases of the software development it is challenging to create accurate performance models as the software system is not yet in production and it is difficult to collect and identify all required empirical information. Especially, it is difficult to get data about the workload and the resource demands of the application. In order to face these challenges, \citet{Barber2004} proposes activities to gather that kind of data:

\begin{itemize}
\item First of all, available production data from existing versions of the software or from external services required for the new system should be analyzed. Using this data the workload, usage scenarios, and resource demands can be extracted.
\item  If no production data is available, design and requirements documents can be analyzed regarding performance expectations for new features. Especially \acp{SLA} can be used as approximation for the performance of external services until measurements are available. 
\item The resource demands can be estimated (\ac{CPU}, \ac{I/O}, etc.) based on the judgments from software developers.
\item After the identification of the available information about the design, workload and resource demands one or more drafts of the performance model should be created and first simulation results should be derived. The results can then be presented and discussed with developers and architects. Then the models should be continuously improved based on further experiments, feedbacks, results, and common sense. 
\end{itemize}

\noindent The challenges of using performance models in early development phases is that it is often difficult to validate the accuracy of the models until a running system exists. Performance predictions based on assumptions, interviews, and pretests can also be inaccurate and subsequently also the decisions based on these predictions. However, a model helps to capture all the collected data in a structured form \citep{brunnert2013integrating}. Considering all these aspects, the following challenges exist for performance evaluations in early development phases that need to be addressed:

\begin{itemize}
\item The trust of the architects and developers in these models can be very low. It is therefore important to make the modeling assumptions and data sources for these models transparent to their users.
\item Design models may be incomplete or inconsistent, especially in the early stages of software development.
\item Modern agile software development processes have the goal to start with the implementation as early as possible and thus may skip the design step. For such methodologies, the approaches outlined in the following section might be better applicable.
\end{itemize}

\subsection{Performance Awareness}
\label{sec:performanceAwareness}

Due to time constraints during software development, non-functional aspects with respect to %
software quality---e.g., performance---are often neglected. %
% However, software developers aim at delivering required functionality regarding certain quality aspects, particularly the performance of a system in terms of response times, resource utilization and throughput. 
Performance testing requires realistic environments, access to test data, and implies the application of specific tools. Continuously evaluating the performance of software artifacts also decreases the productivity of developers. For quality aspects such as code cleanness or bug pattern detection, a number of automatic tools supporting developers exist. Well-known examples are Checkstyle\footnote{http://checkstyle.sourceforge.net} and FindBugs\footnote{http://findbugs.sourceforge.net}. Tools that focus on performance aspects are not yet widely spread, but would be very useful in providing awareness on the performance of software to developers.

Performance awareness describes the availability of insights on the performance of software systems and the ability to act upon them \citep{tuma2014}. \citeauthor{tuma2014} divides the term performance awareness into four different dimensions:
\begin{enumerate}
\item The awareness of performance-relevant mechanisms, such as compiler optimizations, supports  understanding the factors influencing performance. 
\item The awareness of performance expectations aims at providing insights on how well software is expected to perform. 
\item Performance awareness also intends to support developers with insights on the performance of software they are currently developing.
\item Performance-aware applications are intended to dynamically adapt to changing conditions.
\end{enumerate}
The most relevant perspectives for DevOps are the performance awareness of developers and the awareness of performance expectations. Gaining insights into the performance of the code they are currently developing is an increasingly difficult task for developers. Large application system architectures, a continuous iteration between system life cycle phases, and complex \ac{IT} governance represent great challenges in this regard. Complex system of systems architectures often imply a geographical, cultural, organizational, and technical variety. The structure, relationships, and deployment of software components are often not transparent to developers. Due to continuous iterations between development and operations, the performance behavior of components is subject to constant change. The responsibility for components is also distributed across different organizational units, increasing the difficulty to access monitoring data. Additionally, developers require knowledge in performance 
engineering and in using corresponding tools. A number of approaches propose automated and integrated means to overcome these challenges and supporting developers with performance awareness. Selected existing approaches either provide performance measurements \citep{Heger2013, Horky2015, Bures2014} or performance predictions \citep{Weiss2013, Danciu2014} to developers. A brief overview of these approaches is provided below.

Measurement-based approaches collect performance data during unit tests or during runtime. \citet{Heger2013} propose an approach based on measurements collected during the execution of unit tests that integrates performance regression root cause analysis into the development environment. When regressions are detected, the approach supports the developer with information on the change and the methods causing the regression. The performance evolution of the affected method is presented graphically as a function and the methods causing the regression are displayed. % as a graph. %
\citet{Horky2015} suggest enhancements to the documentation of software libraries with information on their performance. The performance of libraries is measured using unit tests. Tests are executed on demand once the developer looks up a specific method for the first time. Tests can be executed locally or on remote machines. Measurements are then cached and refined iteratively. The approach proposed by \citet{Bures2014} integrates performance 
evaluation and awareness methods into different phases of the development process of autonomic component ensembles. High-level performance goals are formulated during the requirement phase. As soon as software artifacts become deployable, the actual performance is measured. Developers receive feedback on the runtime performance within the \ac{IDE}. Measurements are represented graphically as functions within a pop-up window. %
% The actual performance behavior of applications can only be observed during runtime. However, %
At runtime it may be unclear whether the observed behavior also reflects the expected one. Approaches for supporting the awareness of performance expectations provide a means to formulate, communicate, and evaluate these expectations. \citet{Bulej2012} propose the usage of the \ac{SPL} to express performance assumptions for specific methods in a hardware-independent manner. Assumptions on the performance of a method are formulated relative to another method and are not specified in time units. At runtime, assumptions are evaluated and potential violations can be reported to the developer. \citeauthor{Bures2014} employ \ac{SPL} during design to capture performance goals and assign them to individual methods. These assumptions are then tested during runtime.

Model-based prediction approaches aim at supporting developers with insights on the performance of software before it is deployed. The approach by \citet{Weiss2013} evaluates the performance of persistence services based on tailored benchmarks during the implementation phase. The approach enables developers to track the performance impact of changes or to compare different design alternatives. Results are displayed within the \ac{IDE} as numerical values and graphically as bar charts. The approach is only applicable for \ac{JPA} services, but instructions on how to design and apply benchmark applications to other components are also provided by the authors. The approach proposed by \citet{Danciu2014} focuses on the \ac{Java EE} development environment. The approach supports developers with insights on the expected response time of component operations they are currently implementing. Estimations are performed based on the component implementation and the behavior of required services. 

\

We identify the following open challenges and future research directions in the field of performance awareness:
\begin{itemize}
\item Current approaches mainly focus on providing insights on performance but omit enabling developers to act upon them. Future research should investigate how guidance for correcting problems could be provided.
\item Insights compiled for the specific circumstances of developers should be collected and exchanged with Ops. Thus, new trends can be identified and corresponding measures can be taken.
\item The acceptance of performance awareness approaches by developers needs to be evaluated more extensive and improved. Increasing the acceptance will foster the diffusion of these approaches into industry.
\item The performance improvements which can be achieved by employing performance awareness approaches need to be evaluated using industry scenarios.
\end{itemize}
% The approaches presented in this section automate performance evaluation tasks and assure the access to the latest performance measurements.
\subsection{Performance Anti-Pattern Detection}
\label{sec:performanceAntiPattern}
Software design patterns \citep{gamma1994design} provide established template-like %
solutions to common design problems to be used consciously during development. %
By contrast, software performance anti-patterns \citep{smith2000software} constitute design, development, or deployment mistakes with a potential
impact on the software's performance. Hence, anti-patterns are primarily used as a feedback mechanism for different
stakeholders of the software engineering process (e.g., software architects, developers, system operators, etc.). Hereby,
approaches for the detection of performance anti-patterns constitute the basis for anti-pattern-based feedback. 
There are different approaches for performance anti-pattern detection utilizing different detection methodologies,
requiring different types of artifacts and being applicable in different phases of the software engineering process.
Some of these approaches can be integrated into \acp{IDE} and in this way support performance awareness %
(see \autoref{sec:performanceAwareness}) by providing direct feedback on occurring performance mistakes.

\subsubsection{The Essence of Performance Anti-Patterns}
There is a large body of scientific and industrial literature describing different performance anti-patterns 
\citep{smith2000software,smith2002new,smith2002spa,smith2003more,dudney2003j2ee,dugan2002sisyphus,boroday2005dynamic,
understandingHiccups, insufCache, top10PerfProblems, topJavaMemoryProblems, perfAntiPattern2013}. 
All definitions of performance anti-patterns have in common that they describe circumstances that may lead to
performance problems under certain load situations.
However, the definitions of performance anti-patterns conceptually differ with respect to different dimensions.
While some anti-patterns describe mistakes on the architecture level (e.g., Blob anti-pattern \citep{smith2000software}),
others refer to problems on the implementation level (e.g., Spin Wait anti-pattern \citep{boroday2005dynamic}) or even
deployment-related problems (e.g., Unbalanced Processing \citep{smith2002new}).
Furthermore, definitions of anti-patterns differ in the level of abstraction.
While some anti-patterns describe high-level symptoms of performance problems (e.g., The Ramp anti-pattern
\citep{smith2002new}), other anti-patterns describe application-internal indicators or even root causes (e.g., Sisyphus
Database Retrieval \citep{dugan2002sisyphus}).
Furthermore, anti-patterns may describe structural (e.g., Blob anti-pattern \citep{smith2000software}) or behavioral
patterns (Empty Semi Trucks anti-pattern \citep{smith2003more}).
Depending on the types of anti-patterns, different detection approaches are more or less suitable for their detection.

\subsubsection{Detection Approaches}
Approaches for the detection of performance anti-patterns can be divided into %
model-based approaches and measurement-based approaches. Their categories of approaches imply different circumstances under which
they can be applied, yielding different limitations and benefits. In the following, we briefly
discuss the two categories of detection approaches.

\paragraph*{Model-based Approaches}
Model-based approaches for the detection of performance anti-patterns \citep{trubiani2011detection,
cortellessa2007framework, xu2012rule, cortellessa2010process} require %
 architectural (e.g., \ac{PCM} or \ac{MARTE}) or %
 analytic (e.g., \ac{LQN}) %
 performance models, as introduced in \autoref{sec:modelBasedPerformancePrediction}. %
Representing performance anti-patterns as rules (e.g., using predicate logic \citep{trubiani2011detection}) allows to
capture structural as well as behavioral aspects of performance anti-patterns. While certain structural and behavioral %
aspects can be evaluated
directly on the architectural model, associated performance-relevant runtime aspects can be derived %
by performance model analysis or simulation. Applying anti-pattern detection rules to the models allows to identify flaws in the
architectural design that may impair software performance. Due to the abstraction level of architectural
models, the detection scope of model-based approaches is inherently limited to architecture-level anti-patterns. In
particular, performance anti-patterns that are manifested in the details of implementation cannot be detected by
model-based approaches. Furthermore, due to the high dependency on the models, the detection accuracy of model-based
anti-pattern detection approaches is tightly coupled to the quality (i.e., accuracy and representativeness) of the
architectural models.

\paragraph*{Measurement-Based Approaches} %

Depending on their stage of usage in the software lifecycle, %
measurement-based approaches can be further divided into test-based %
and operation-time anti-pattern detection approaches: %

{Test-based anti-pattern detection approaches}
utilize performance tests (e.g.,
as part of integration testing \citep{joergsen1994integration}) to gather performance measurement data as basis for
further reasoning on existing performance problems %
\citep{Wert2013, Wert2014, grechanik2012automatically}. Thereby, measurement-based approaches (see 
\autoref{sec:performanceMeasurement}) are applied to retrieve performance data of interest.
As monitoring tools introduce measurement overhead that may impair the accuracy of measurement data, systematic
experimentation \citep{westermann2013deriving, Wert2013} can be applied to deal with the trade-off between accurate
measurement data and high-level of detail of the data. Similarly to the model-based
approaches, test-based detection approaches apply analysis rules that evaluate the measurement data to identify
potential performance anti-patterns.
As test-based detection approaches rely on execution of the system under test, a testing environment is required that is
representative to the actual production environment. In order to save costs, the testing
environment is often considerably smaller than the production environment. As detection of performance anti-patterns is often
relative to the performance requirements, in these cases, performance requirements need to be scaled down to the size of
the testing environment in order to allow reasonable, test-based detection of performance anti-patterns.
Furthermore, test-based detection approaches utilize load scripts for load generation during execution of performance
tests. Hence, the detection accuracy highly depends on the quality (i.e., representativeness of real users) of the load
scripts. The DevOps paradigm is the key enabler to derive representative load scripts for test-based anti-pattern
detection from production workloads (\autoref{sec:workloadCharacterization}). %
Test-based approaches analyze the implemented target system in its full level of detail. They potentially cover all
types of anti-patterns: from architecture-level via implementation-level through to structural as well as behavioral anti-patterns.
Finally, test-based approaches that run fully automatic \citep{Wert2013, Wert2014} can be assimilated into \ac{CI} \citep{duvall2007continuous} in order to provide frequent, regular feedback on potential existence of
performance anti-patterns in the code of the target application.

{Operation-time anti-pattern detection approaches,} e.g., by \citet{parsons2004framework}, are similar to
test-based approaches with respect to the detection methodology. However, as they are applied on production system
environments, they entail additional limitations as well as benefits. In a production environment, the measurement overhead induced by monitoring tools is a much more critical factor than with test-based approaches, as the
monitoring overhead must not noticeably affect the performance of real user requests. Therefore, operation-time
anti-pattern detection approaches apply rather coarse-grained monitoring, which affects the ability of providing
detailed insights on specific root causes of performance problems. Furthermore, performance anti-patterns that are
detected by operation-time approaches might already have resulted in a performance problem experienced by end users.
Hence, operation-time detection of performance anti-patterns is rather reactive. However, as performance characteristics
are investigated on the real system, under real load, operation-time approaches are potentially more accurate than
model-based or test-based approaches.

\

We see the following research challenges in the area of performance anti-pattern detection: %
\begin{itemize}
 \item Anti-patterns are usually described in textual format. More work is %
       needed to formalize these descriptions into machine-processable rules %
       and algorithms. %
 \item Many approaches use fixed thresholds in their detection rules and algorithms, %
       e.g., in order to judge whether a number of remote communications is %
       indicative for a performance problem. %
       This leads to context-specific and system-specific configurations. %
       More research is needed to automatically determine suitable thresholds or %
       to completely avoid them. %
 \item More research is also needed to better understand and formalize the %
       relationship between symptoms, indicators, and root-causes connected %
       to performance anti-patterns and performance problems in general. %
\end{itemize}

\subsection{Performance Change Detection}
\label{sec:performanceChangeDetection}
A key goal of a tighter integration between development and operations teams is to better cope with changes in the business environment. In order to react quickly in such situations, a high release frequency is necessary. This changes the typical software release process in a way that, instead of releasing new features or bug fixes in larger batches in a few major versions, they are released more frequently in many minor releases. 

The performance characteristics of \acp{EA} can change whenever new features or bug fixes are introduced in new versions. Due to this reason, it is necessary to continuously evaluate the performance of \ac{EA} versions to detect performance changes before an \ac{EA} version is moved to production. % \citep{Brunnert:2014:VALUETOOLS}. %
The previously introduced approaches during design and development cannot capture all performance-related changes. Activities during the design phase might not capture such changes because minor bug fixes or feature additions do not change the design and are thus not detectable. During the implementation phase, only changes are detectable that are caused by the code of an \ac{EA} directly. Therefore, changes that are only detectable on different hardware environments (e.g., in different deployment topologies) or in specific workload scenarios must be analyzed before an \ac{EA} version is released.

In order to analyze such performance-related changes of \ac{EA} deployments on different hardware environments or for multiple workload scenarios, measurement- and model-based performance evaluation techniques can be used. Measuring the performance of each \ac{EA} version is often not feasible because maintaining appropriate test environments for all possible hardware environments and workloads is associated with a lot of cost and effort. Therefore, a mixture of model- and measurement-based performance evaluation approaches to realize performance change detection techniques are introduced in the following.

A measurement-based technique that can be used to detect performance changes on a low level of detail are performance unit tests \citep{Horky:2015:PERFUNITTESTS}. The key idea of such performance unit tests is to ensure that regressions introduced by developer check-ins are quickly discovered. One possible implementation of performance unit testing is to use existing \ac{APM} solutions during the functional unit tests. %
%\footnote{\url{https://community.compuwareapm.com/community/display/PUB/dynaTrace+in+Continuous+Integration+-+The+Big+Picture}}
In order to make the developers aware of any performance regressions introduced by their changes in new \ac{EA} versions, an integration of such tests in \ac{CI} systems % (e.g., Jenkins\footnote{\url{http://jenkins-ci.org/}}) %
is often proposed \citep{Waller:2015:SEN}. %
A measurement-based approach that requires test environments that are comparable to the final production systems is proposed by \citet{Bulej:2005:RRA:1077906.1077924}. The authors propose to use application-specific benchmarks to test the performance for each release. Using such benchmarks has the advantage of repeatability and also makes the results more robust compared to single performance tests.
Another measurement-based approach to detect performance regressions during development is proposed by \citet{Nguyen:2012:ADP:2188286.2188344}. The authors propose to use so-called control charts in order to detect performance changes between two performance tests. Similar to this approach are the works by  \citet{Cherkasova2008,Cherkasova:2009:AAD:1629087.1629089}  and \citet{Mi2008}. The authors propose the use of so-called application signatures. Application signatures describe the response time of specific transactions relative to the resource utilization. However, application signatures are intended to find performance changes for systems that are already in production. 

\begin{figure}
	\centering
	\includegraphics[width=0.8\textwidth]{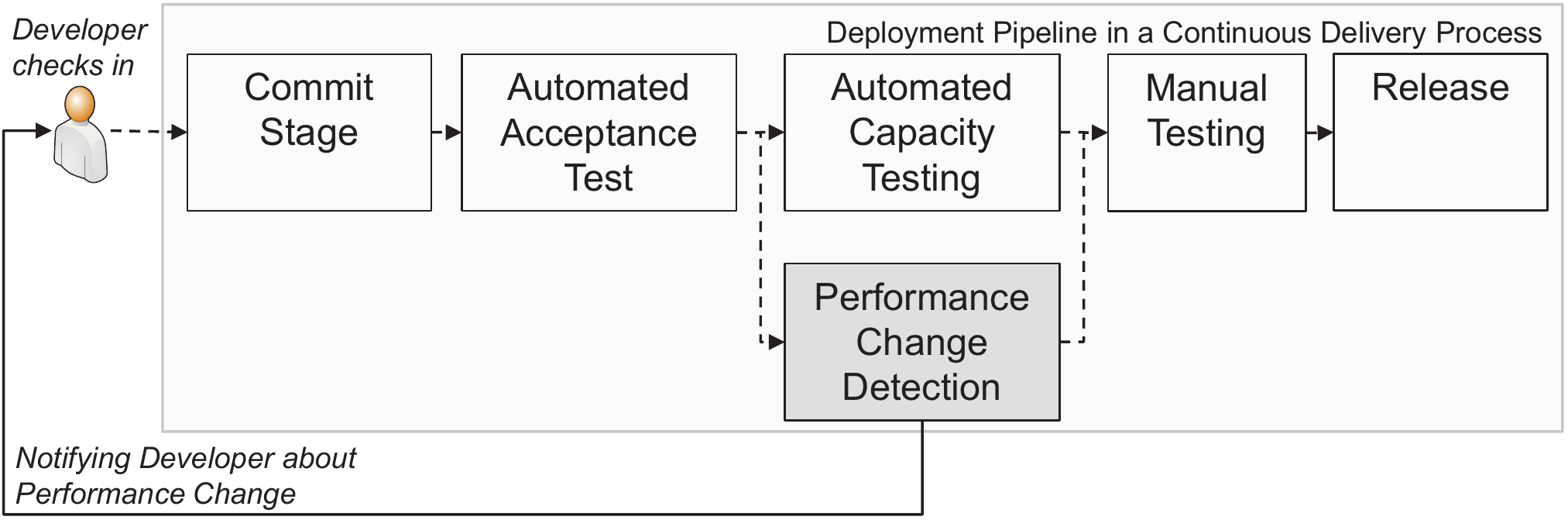}\hfill
	\caption{Detecting performance change in a deployment pipeline \citep{Brunnert:2014:VALUETOOLS}}
	\label{fig:performancechangedetection}
\end{figure}

% As creating and building deployable \ac{EA} versions is only the first step in getting an \ac{EA} in a releasable state, a new paradigm called \acl{CD} emerged in recent years \citep{Humble:2010:CDR:1869904}. The idea of \acl{CD} is to automate all steps from creating a deployable \ac{EA} version until a releasable version exists in a so-called deployment pipeline (see \autoref{fig:performancechangedetection}). %
A model-based performance change detection process within a deployment pipeline, %
depicted in \autoref{fig:performancechangedetection}, %
is proposed by \citet{Brunnert:2014:VALUETOOLS}. This approach uses monitoring data collected during automated acceptance tests in order to create models (called resource profiles). These resource profiles describe the resource demand for each transaction of an \ac{EA} and are managed in a versioning repository in order to be able to access the resource profiles of previous builds. The resource profile of the current \ac{EA} version is used to predict performance for predefined hardware environments and workloads. The 
prediction is performed with predictions derived from resource profiles of one or several previous versions. The prediction results are compared with each other used as an indicator for change. The resource profiles themselves and the check-ins that triggered a build are then analyzed in order to find the source of a change (e.g., by comparing two resource profiles). %

Compared to several works that introduce approaches to detect performance change, there is a relatively low amount of work on identifying the reasons for a performance change. One of the few examples is the work by \citet{Sambasivan:2011:DPC:1972457.1972463}. The authors propose an approach to analyze the reasons of a change based on request execution flow traces. Their approach is based on the response time behavior of single component operations involved in the request processing and their control flow. It might be interesting to combine this approach with the approach presented by  \citet{Brunnert:2014:VALUETOOLS} because the resource profiles and their prediction results provide all the necessary data for the root cause search approach presented by \citet{Sambasivan:2011:DPC:1972457.1972463}. This would allow to detect and analyze performance change in a completely model-driven way.

Even though there is a lot of work going on in the area of performance change detection, a lot of open challenges remain, such as:

\begin{itemize}
\item Test coverage is always limited: performance changes in areas that are not covered by a test workload cannot be detected.
\item Model-based performance change detection techniques are always associated with the risk that aspects with impact on performance are not properly reflected in a model. 
\item A lot of the existing model-based performance evaluation techniques focus on \ac{CPU} demand, if memory, network, or \ac{HDD} would cause a performance regression, it cannot be detected using such models.
\item Test coverage is not only limited for software itself but also in terms of the amount of workloads and hardware environments that can be tested in a reasonable time frame.
\end{itemize}

\section{Application Performance Management During Operations} % (Ops)
\label{sec:integrationPhase}

Once an \ac{EA} is running in a production environment it is important to continuously ensure that it meets its performance goals. The activities required for this purpose are summarized by the term \ac{APM}. \ac{APM} activities are required regardless how well \ac{SPE} activities during development outlined in the previous section have been executed. This is the case because either assumptions about the production environment or the workload can be wrong. Furthermore, performance data collected during operations provides a lot of insights for the development teams to get them from assumptions to knowledge. The key \ac{APM} activities during development are outlined in this section as follows: \autoref{sec:performanceMonitoring} outlines one of the most fundamental activities, namely performance monitoring. Afterwards, \autoref{sec:problemdetectionanddiagnosis} covers performance problem detection and diagnosis activities based on data collected using monitoring techniques. In order to reduce the 
need for  manual interaction, the section concludes in %
\autoref{sec:modelsatruntime} with existing approaches and challenges regarding the application of performance models to control the performance behavior of an \ac{EA} autonomously.

\subsection{Performance Monitoring}
\label{sec:performanceMonitoring}
\newcommand{\localTODO}[1]{} % #1

\citet{Gartner2014APMQuad} use the following five dimensions %
of \ac{APM} functionality to assess the commercial market of %
\ac{APM} tools in their yearly report: %
\begin{inlineenumerate}
 \item \ac{EUM}, %
 \item application topology discovery and visualization, %
 \item user-defined transaction profiling, %
 \item application component deep dive, and %
 \item \ac{ITOA}. %
\end{inlineenumerate}
% In the remainder of section, we will use this classification %
% to give an overview about performance monitoring activities. %

The \ac{EUM} dimension stresses the need to include the %
monitoring of client-side performance measures---%
particularly end-to-end response times---instead of looking %
at only those performance measures obtained inside the server %
boundaries. %
Major reasons for this requirement are that, nowadays, %
\begin{inlineenumerate}
\item \acp{EA} move considerable parts of the request %
processing to the client, e.g., rendering of \acp{UI} %
in web browsers on various types of devices; and that %
\item networks are increasingly influencing the user-perceived %
performance because \acp{EA} are accessed via different types %
of connections (particularly mobile) and from %
locations all over the world. %
\end{inlineenumerate}
\ac{EUM} is usually achieved by adding instrumentation %
to the scripts executed on the client side and sending %
back the performance measurements as part of subsequent %
interactions with the server. %

Application topology discovery and visualization comprises %
the ability to automatically detect and present
information about the components and %
relationships of \ac{EA} landscapes % 
as well as to make this information analyzable. %
An \ac{EA} landscape consists of different types of physical %
or virtual servers hosting software components that %
interact with each other and with third-party services  via different integration %
technologies, including synchronous remote calls and %
asynchronous message passing. %
Application topologies are usually discovered by %
monitoring agents deployed on the application servers, %
which send the obtained data to a central monitoring %
database. Topologies are usually represented and %
visualized as navigable graph-based representations %
depicting the system components and control flow. %
These graphs are enriched by aggregated performance data %
such as calling frequencies, latencies, response times, %
and success rates of transactions, as well as %
utilization of hardware resources. %
This information enables operators to get an overall %
picture of a system's health state, particularly with %
respect to its performance, and to detect and diagnose %
performance problems. %

User-defined transaction profiling refers to the %
functionality of mapping implementation-level details %
about executed transactions (e.g., involved classes %
and methods, as well as their performance properties) %
to their corresponding business transactions. This %
feature is useful in order to assess the impact of %
performance properties and problems to business %
indicators. For example, it can be evaluated which %
business functions, such as order processing, are %
affected in case certain software components are %
slow or even unavailable. %

Application component deep dive refers to the %
detailed tracing and presentation of %implementation-level %
call trees for %each processed %
transactions. The call trees include %
control flow information, including executed software %
methods, remote calls to third-party services, and %
exceptions that were thrown. %
The call tree structure is enriched by %
performance measurements, such as response %
times and execution times, as well as resource %
demands. %

Orthogonal to the previous dimensions, \ac{ITOA} %
functionality aims to derive higher-order information %
from the data gathered by the first four dimensions, %
e.g., by employing statistical analysis techniques, %
including data mining. %
%    ``complex operations event processing, %
%    statistical pattern discovery and recognition, %
%    unstructured text indexing, search and inference, %
%    topological analysis, %
%    multidimensional database search and analysis'' %
A typical example for \ac{ITOA} is performance %
problem detection and diagnosis as presented in %
\autoref{sec:problemdetectionanddiagnosis}. %

The following list includes a summary of selected challenges %
regarding the current state of performance monitoring with a %
focus on the \ac{APM} tooling infrastructure: %

% \todo{Refine challenges}

\begin{itemize}
 \item The most mature \ac{APM} tools are closed-source software products %
       provided by commercial vendors. %
       These tools provide comprehensive support for monitoring heterogeneous %
       \ac{EA} landscapes, covering instrumentation support for various %
       technologies and including novel features such as adaptive instrumentation %
       and automatic tuning for reduced performance overhead. %
       However, being closed source, the tools' functionality often cannot be 
       extended or reused for other purposes in external tools. %
       Also, details about the functionality are generally not published. %
       Moreover, researchers are usually not allowed to evaluate or compare their %
       research results with the capabilities of the tools as this is %
       not permitted according to the license agreements. %
 \item The data collection functionalities of \ac{APM} tools---including %
       data about \ac{EA} topologies, transaction traces, and performance %
       measures---are a valuable input for performance model extraction %
       approaches, as presented \autoref{sec:performanceModelExtraction}. %
       However, all too often, no defined interfaces exist to access this %
       data from the tools. In order to increase the interoperability of %
       \ac{APM} platforms, open or even common interfaces are desirable. %
       In this way, researchers can particularly contribute to the effectiveness %
       of \ac{APM} solutions by developing novel (model-based) \ac{ITOA} approaches %
       that build on the data collected by the mature \ac{APM} tools. %
%  \item One key aspect of software systems is their observation at runtime %
%        to be able to evaluate runtime qualities, such as performance.
%        While with present monitoring framework this task is highly technology %
%        dependent resulting in different specific solutions for each technology
%        resulting in problems for the integration of measurements.%
%        For example, Kieker \citep{vanhoorn2014dynamic} supports various %
%        languages including C, Perl and Java.
%        For each language and each framework implemented in those languages separate probes and samplers have been implemented and must be configured and setup separately.%
 \item The configuration of \ac{APM} tools, due to the system-specificity, is very %
        complex, time-consuming, and error prone---particularly given the %
        aforementioned faster release and deployment cycles requiring a %
        continuous refinement.
        For example, \acs{APM}-specific questions such as what and where to monitor %
        are decisions mainly taken by operations teams. However, %
        performance models used during design time could be further exploited %
        and extended to specify operations-related monitoring aspects earlier %
        and more systematically in the system lifecycle.  
        An automatic configuration of \ac{APM} tools is desirable, e.g., based on %
        higher level and tool-agnostic descriptions of monitoring goals attached %
        to architectural performance models. %
%         Existing model-based and model-driven solutions try to provide a unified %
%         method to specify probe and probe placements. However, the comparability %
%         and the ability of integrating the results is still an open issue, as %
%         well as the development of generators to map the model to specific %
%         implementations. %
\end{itemize}

%!TEX root = ../maindoc.tex
\subsection{Problem Detection and Diagnosis}
\label{sec:problemdetectionanddiagnosis}

% This section serves to summarize the state-of-the art in APM research and %
% practice---focusing on the detection and diagnosis of performance problems---%
% and to identify current limitations and research challenges. %

% \begin{itemize}
% 	\item Detection: Determine that a symptom is present, e.g., increased/decreased 
% 	      response times, resource utilization, number of errors
% 	\item Diagnosis: Determine the root cause(s) of a symptom
% \end{itemize}

Performance problem detection aims to reveal symptoms for present %
or upcoming system states with degraded or suspicious  performance properties, %
unusually high or low response times, utilization of resources, or %
number of errors. %
The diagnosis step aims to reveal the root cause of the performance problems %
observed by the previously detected symptom(s). These steps are comparable to the activities during anti-pattern detection outlined in \autoref{sec:performanceAntiPattern}. However, their main difference is that problems revealed by anti-pattern detection approaches are limited to scenarios that are known to cause problems, whereas general problem detection tries to reveal problematic situations without prior knowledge. Approaches for problem detection and diagnosis can be classified based on %
various dimensions. %
In this section, we focus on %
 \emph{when} the analysis is conducted (before or after the problem occurs), %
 \emph{who} is performing it (a human or a machine), and %
 \emph{how} it is performed (based on information about the system state or individual transactions). %
Note that we are not aiming for a complete %
taxonomy and/or classification of approaches but we give examples for how %
performance problem detection and diagnosis can be conducted and what example %
approaches are. %

\subparagraph*{When?}        
Reactive approaches aim to detect and diagnose problems after they occurred, %
using statistical techniques like detection of threshold violations or %
deviations from previously observed baselines. %
Proactive approaches aim to detect and/or diagnose problems before %
they occur \citep{SalfnerLenkMalek2010ASurveyOfOnlineFailurePredictionMethods}, %
using forecasts/predictions based on historic data for %
performance measures of the same system. %
Forecasting and prediction techniques include mature statistical %
techniques like time series forecasting, machine learning, as well as %
a combination of these techniques with model-based performance %
prediction incorporating architectural knowledge about the architecture %
(e.g., as in the approach by \citet{RathfelderBeckerKrogmannReussner2012WorkloadAwareSystemMonitoringUsingPerformancePredictionsAppliedToALargeScaleEMailSystem}). %
%\emph{Who: automatic vs.\ manual.} %
%Approaches that include a human-in-the loop are called manual. %

\subparagraph*{Who?}   
Problem detection is usually achieved by setting %
and controlling baselines on performance measures of interest, %
both of which may be conducted manually or automatically. %
Another approach is anomaly detection \citep{ChandolaBanerjeeKumar2009AnomalyDetectionASurvey,CSMR2009,ICAC2011}, %
which aims to detect patterns in the runtime behavior that %
deviate from previously observed behavior. %
If no automatic problem detection is in place, problems will be %
often reported by end users or by system operators that inspect %
the current health state of the system. %
Manual diagnosis of problems is usually performed by inspecting %
monitored data or by reproducing and analyzing the problem in %
a controlled development environment, using tools like debuggers %
and profilers. %
In any case, expert knowledge about typical relationships between %
symptom and root cause %
can be used to guide the diagnosis strategy (e.g., based on the %
performance anti-patterns presented in \autoref{sec:performanceAntiPattern}). 

\subparagraph*{How?}  
%\emph{How: state-based vs.\ transaction-based.} %
State-based problem detection and diagnosis approaches reason %
about aggregate behavior of system or component measures obtained %
from a certain observation period, e.g., %
response time percentiles or distributions, and total invocation %
counts. Note that data from individual transactions (e.g., individual %
response times and control flow) may be used for the aggregation %
and model generation, but are dropped after the aggregation step (e.g., as in the approach of \citet{AgarwalApplebyGuptaKarNeogiSailer2004ProblemDeterminationUsingDependencyGraphsAndRunTimeBehavior}). %
Transaction-based approaches (e.g., the approach of \citet{Kiciman2005DetectingApplicationLevelFailuresInComponentBasedInternetServices}) %
are usually triggered by symptoms of %
performance problems observed for (a class of) transactions, e.g., %
high response times for a certain business transaction or error %
return codes. For diagnosis, the transaction's call tree is inspected %
similar to the process of profiling, e.g., looking for methods with %
(exceptionally) high response times, high frequencies of method %
invocations, or exceptions thrown. %

\

The following list includes a summary of selected limitations of %
current approaches and research challenges: %

\begin{itemize}
 \item Performance requirements, e.g., \acsp{SLA} for response times, are often barely %
       defined in practice. This makes the intuitive approach of %
       comparing (current or predicted) performance measures with thresholds %
       or baselines infeasible. Approaches are needed to automatically %
       derive meaningful baselines from historic measurements. %
       Note that a feasible trade-off of classical classification quality %
       attributes such as false/true positives/negatives needs to be achieved %
       to let administrators trust the performance problem detection and diagnosis %
       solution. %
 \item Faster development and deployment cycles, in addition to potentially %
       multiple components deployed in different versions at the same time, impose %
       challenges to configurations of problem detection and diagnosis approaches. %
%        e.g., the afore-mentioned baselines become obsolete very fast. %
 \item Basic problem detection support, often
       based %
       on automatically determined baselines, is provided by some commercial %
       \ac{APM} tools. %
       However, for researchers, it is often hard to judge the underlying %
       concepts, because the tools are closed source (exceptions exist) %
       and the concepts are protected by patents or not published at all. %
       In order to improve the comparability and interoperability of %
       tools, open \ac{APM} platforms are desirable. %
%  \item System perturbation (trade-off between detail and overhead/disturbance)
%  \item Configuration of monitoring tools
%  \item Configuration of problem detection/diagnosis (thresholds, baselines)
%  \item Configuration of rules set
%  \item ML-based approaches required a large training set but failures are rare and %
%        the model may be outdated.
 \item Basic automatic problem detection is accepted by administrators, given %
       that the false/true positive/negative rates are in an acceptable %
       range. As fully automatic problem resolution is usually not %
       accepted, future work in the performance community could be to %
       focus more on recommender systems for problem diagnosis and resolution, %
       which can be based on expert knowledge. %
\end{itemize}
\subsection{Models at Runtime}
\label{sec:modelsatruntime}

Self-adaptive or autonomic software systems use models at runtime 
\citep{SalehieTahvildari2009SelfAdaptiveSoftwareLandscapeAndResearchChallenges}, %
which continuously perform activities in a control loop, as follows: %
\begin{inlineenumerate}
% 	\item evaluating changes in the \ac{EA} topology (either in the software architecture or deployment configuration) of the application;
	\item updating the model with monitoring data by integrating with appropriate monitoring facilities; % \citep{Balduini2013};
	\item learning, tuning, and adjusting model parameters by adopting %
	      appropriate self-learning techniques;
    \item employing the model to reason about adaptation, scaling, %
              reconfiguration, repair, and other change decisions. %
\end{inlineenumerate}
Several frameworks have been developed to implement runtime engines for these %
activities. Recent examples include 
       EUREMA \citep{Vogel2014}, 
       iObserve \citep{MRT2014}, %
       MORISIA \citep{Vogel2011}, %
       SLAstic \citep{vanHoorn14PhDThesis}, and %
       Zanshin \citep{Tallabaci2013}. %
%These frameworks %, however, 
%use different modeling formalisms and reasoning techniques for %
%evaluating and verifying system properties. % For instance, Rainbow \citep{Garlan2004} %
%uses architectural models, while Zanshin \citep{Tallabaci2013} uses goal models. %
%As another example, MUSIC uses rule-based technique for reasoning, while %
%\citet{Filieri2012a} use control theory for verifying system properties. %
%Focusing on performance, SLAstic employs \ac{PCM}-like architectural %
%models with performance annotations. %

\begin{figure}
	\centering
	\includegraphics[width=0.7\textwidth]{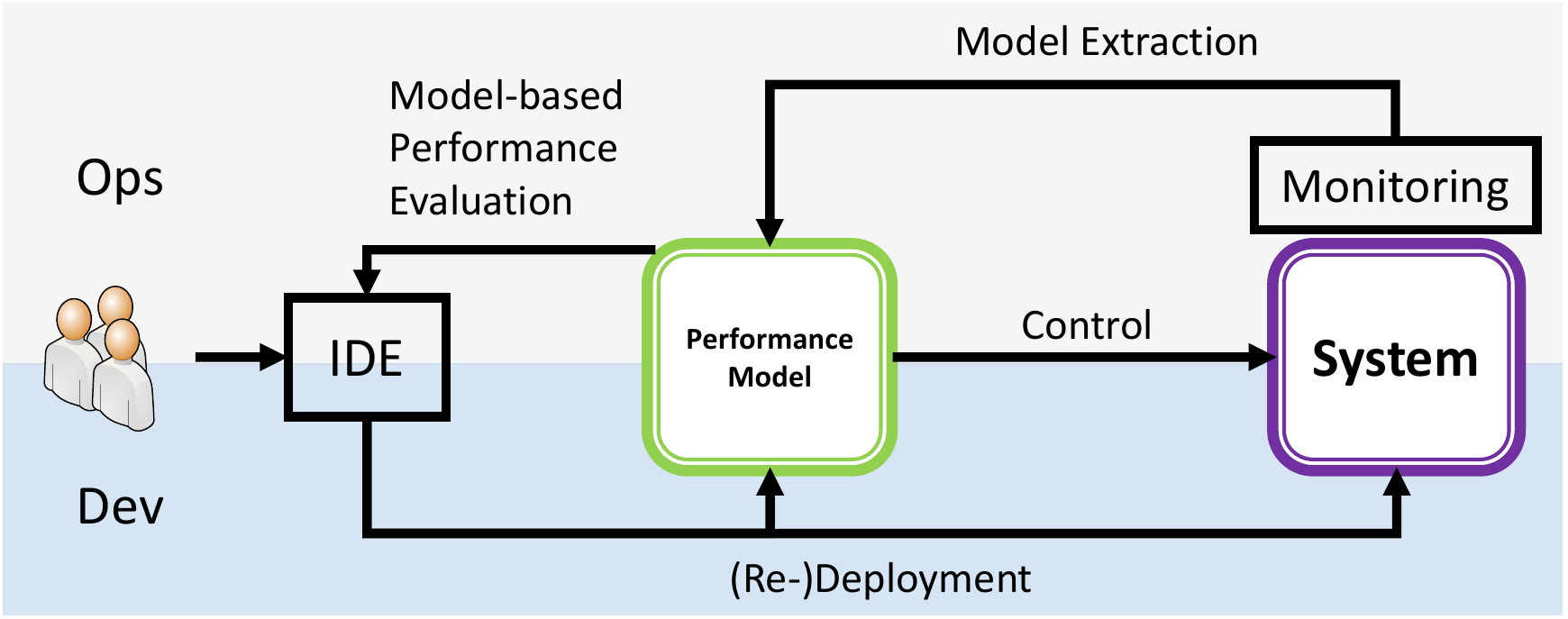}\hfill
	\caption{Models at runtime in DevOps}
	\label{fig:mrt}
\end{figure}

Both development and operations can benefit from models at runtime in %
the context of DevOps, as illustrated in \autoref{fig:mrt}. %
Initially, performance-augmented models and implementation artifacts %
are deployed to operations. 
% an \ac{IDE} is used to design analytical models representing certain %
% aspect of the system and also can be used to provide the application %
% performance analysis information to the developer. %
The parameterized models at runtime are continuously updated and become more %
accurate by receiving runtime monitoring data. %
As mentioned above, the performance models %
may serve as a basis to dynamically control a system's performance during %
operation. %
For instance, the updated models can be used for runtime capacity management %
(\autoref{sec:capacityPlanning}) helping the operations team to decide about %
the resources for the system or providing feedback for the auto-scaling %
mechanism.
Models at %
runtime can have several purposes for operations teams. They can be exploited as the source for monitoring aspects of a running system, to affect the system via model manipulation, and as a basis for analytical methods, such as model-based verification and model-based simulation. %
% Models are in general useful for analysis purposes as they provide high-level knowledge about the system. The high-level knowledge is easy to interpret by automated analytical techniques and can be used for feeding back the Dev process or closing the loop around it for enabling self-adaptive mechanisms.
A promising role of models at runtime for development  in the context of DevOps is to bring back %
the adapted model along with the associated runtime performance information to %
the development environment. The gathered information can be used to analyze and improve %
software system performance based on refined design-time artifacts such as software %
architectural descriptions, employing the model-based evaluation techniques %
summarized in this report. %
Using the updated models, %
developers can then update the system design and the underlying code or %
it can be automatically updated by appropriate techniques that causally %
connect the models and the system \citep{Chauvel2013} in order to improve %
system performance.
Note that all major components in \autoref{fig:mrt} reside in both \ac{Dev} and \ac{Ops} since the system itself has both design-time and runtime aspects and this is the same for models and the 
development and operational team.

% More concretely, in the scope of DevOps, models at runtime can be used as a %
% basis for verifying performance properties of the system during Ops and feeding %
% back vital information for Dev to improve design-time artifacts. %
% From this perspective, models can play various roles \citep{Bencomo2014}. %
% Depending on what the models represent, they can be adopted as a primary source %
% of information to reason about different aspects of the running system. %
% For instance, performance models can be used to 
% \begin{inlineenumerate}
% \item represent the performance constraints to be satisfied, %
% \item the current state of the system, %
% \item performance prediction of the system, and %
% \item the steady state can be calculated based on them.  
% \end{inlineenumerate}
% The modeling approaches that are used at design-time enable the verification of certain properties of the system under study. The use of models at runtime, however, has the advantage that some of the constraints are relaxed as there available more knowledge at runtime, for example the current runtime state is available for reasoning, reaction, and regulation. At development time, verification is
% required to reason about all possible states. Several of the variables that are unknown
% at development time become known at runtime and can allow for a more focused analysis
% of the current state and possibly several neighboring ones. A running system can continually monitor these aspects and react to them. 

\

Selected challenges regarding models at runtime in the context of DevOps %
include: %
\begin{itemize}
	\item Monitoring sensors are not precise and contain noise. As a result, the parameters of the models that are required to be estimated by such monitoring data also get influenced by such measurement inaccuracies. One of the relevant challenges in such context is to develop reliable and robust estimation techniques that can update model parameters accurately given such inaccuracies in measurements.
	\item Monitoring data is collected on an implementation level which might deviate from the model level in various degrees. For example, a monitoring record may contain the signature of the invoked operation, class, and object which must then be mapped to the corresponding component instance or type on model level. Such mapping becomes even more complicated when non-structural properties are observed. In many approaches this mapping is performed by a function that evaluates the signatures and maps them based on an algorithm to a model constructed at runtime. However, in context of pre-existing design-time models, such automatically derived runtime models may not correspond to their design-time counterparts. Therefore, design-time models in conjunction with a mapping between code and model must be available at runtime to provide a correspondence of the monitoring data to model elements. 
% 	\item When the system is updated dramatically, the models that are used for reasoning need to be updated or to be redesigned. Such model update most of the time requires human knowledge and cannot be automated typically. 
% 	\item Automating the process of performance assessment and architecture enhancement with an integrated toolchain is also a relevant challenge in this context.
	\item To be able to feedback knowledge gathered during monitoring, either the runtime model must use a common meta-model with the design-time models or provide an accurate mapping between both models. In case of a common subset it is important to understand the information flow from runtime to design-time and vice versa. For example, a changed deployment at runtime must be reflected in the model. 
	\item Runtime models, like user behavior models, are constructed out of monitoring data which are then used to predict future user and system behavior. However, certain events, such as a sales event for a web-shop application, cannot be predicted correctly out of observed data. Therefore, it must be possible to feed in new user behaviors at runtime without affecting the predicted behavior based on observed behavior.  
\end{itemize}

\section{Evolution: Going Back-and-Forth between Development and Operations}
\label{sec:evolutionPhase}

After a system has been initially designed, implemented, and deployed, the evolution phase of software development starts. As phrased by Lehman's first law of software evolution, a system ``must be continually adapted or it becomes progressively less satisfactory in use'' \citep{lehman2001a}, and thus evolution and change are inevitable for a successful software system. 
In the evolution phase, development and operations can be intertwined as shown in \autoref{fig:ApplicationLifeCycle}. 
While the system is %(hopefully) %
operated continuously, development activities after initial deployment are triggered by specific change events.
With respect to performance, two types of triggers are most relevant:
\begin{itemize}
	\item Changing requirements: Changed functional requirements need to be incorporated in the software architecture, software design, and implementation. Such new or changed functional requirements can newly arise in the form of new feature requests. Alternatively, these new requirements can already be on the release plan for some time before they get tackled in a next development iteration. In addition to functional requirements, new or changed quality requirements can also trigger development. Examples include requirements for better response times as the users expectations have become higher over time (as also discussed by \citet{lehman2001a}). 
	\item Changing environment: In addition to changing requirements, also changes of the environment may create a need to update a system's design. For performance, the most important type of environmental change is a changing workload, both in terms of changing number of users and in terms of changing usage profile per user. A second common type of change concerns the execution environment, such as migration from on-premise to cloud. Such environmental changes, if not properly addressed, can lead to either violating performance requirements (in case of increasing workload) or inefficient operation of the system (in case of decreasing workload). In addition to workload and execution environment, changes in the quality properties of services that the \ac{SUA} depends on can likewise cause the performance of the \ac{SUA} to change. 
\end{itemize}

\noindent As a basis for performance-relevant decision making in the %
evolution phase, runtime information from the system's production %
environment can be exploited using model-based performance evaluation %
techniques. As detailed in previous sections, this includes %
information about the system structure and behavior, as well as workload %
characteristics. %
This section focuses on two specific performance engineering activities %
within the evolution phase, namely %
 capacity planning and management (\autoref{sec:capacityPlanning}), %
 as well as %
 software architecture optimization for performance (\autoref{sec:architecturalImprovements}). %
% The goal of capacity planning and management is to continually %
% provision an adequate configuration of the software and hardware %
% deployment which at the same time satisfy \acp{SLA} and respect cost %
% constraints. %
% Well-known optimization techniques can be used in the performance %
% engineering context to automatically search for adequate software %
% architectural configurations in the usually huge space of available %
% architectural alternatives. %

\subsection{Capacity Planning and Management}
\label{sec:capacityPlanning}
Whenever an \ac{EA} is being moved from development to operations, it is necessary to estimate the required capacity (i.e., the amount and type of software and hardware resources) for given workload scenarios and performance requirements.  This is extremely important for completely new deployments, but it is also required whenever major changes in the feature set or workload of an application are expected. 
In case a new deployment needs to be planned, this activity is usually referred to as capacity planning. As soon as a deployment exists and its capacity needs to be adapted for changing workloads or performance requirements, this activity is referred to as capacity management.

In order to plan capacity, it is important to not only consider performance goals but other perspectives %
such as costs. The latter includes investments in hardware infrastructure and software licenses, %
as well as operations expenditures, e.g., for system maintenance and energy. %
According to \citet{menasce2001}, capacity is considered adequately, if the performance goals are met within given cost constraints (initial and long term cost), and if the proposed deployment topology fits within the technology standards of a corporation. Therefore, estimating the required capacity for a deployment requires the creation of a workload model, a performance model, and a cost model.

Nowadays, a key challenge that leads to the importance of capacity management for existing deployments is that it is often practically not feasible to evaluate the performance of all deployments of an \ac{EA} during development as shown in  \autoref{sec:designPhase}. Therefore, operations teams cannot expect that all their specific scenarios have been evaluated. A challenge for new deployments is, that not all deployments are known at the time of a release. Furthermore, there is often a lack of information (e.g., about the resource demands of an application for specific transactions) whenever a new deployment needs to be planned. 

The traditional way of approaching capacity planning and management activities is to setup a test environment, execute performance tests, and use the test results as input for capacity estimations \citep{king2004performance}. As the test environments for such tests need to be comparable to the final production deployments, this approach is associated with a lot of cost and manual labor. Therefore, model-based performance evaluation techniques are proposed in research results in order to reduce these upfront investment costs \citep{menasce2001}.

One example for a model-based capacity planning tool is proposed by \citet{liu2004capacity}. This tool can be used to support capacity planning for business process integration middleware. A similar tool for component- and web service-based applications is proposed by \citet{Zhu2007}. However, their tool is intended to be used to derive capacity estimations from designs and these estimations cannot be used for final capacity planning purposes.
%
% A model-based approach to support capacity planning and management activities is proposed by %
\citet{Brunnert:2014:UAP:2602576.2602587} use resource profiles that serve as an information sharing entity between the different parties involved in the capacity planning process. Resource profiles can be complemented with workload and hardware environment models to derive performance predictions.

As all of the aforementioned approaches require a manual interaction to configure a model in a way that performance, cost, and technological constrains are met, an automated optimization is proposed by \citet{li2010sla}. The authors propose an automated sizing methodology for Enterprise Resource Planning systems that takes hardware and energy costs into account. This methodology tries to automatically find a deployment topology which provides adequate capacity for the lowest total cost of ownership.

In addition to the previously mentioned capacity planning and management activities %
usually performed offline with a longer time horizon, a lot of work is currently done  %
in the area of self-adaptation and runtime resource management \citep{Kounev2011,vanHoorn14PhDThesis}. % dup: WUP2009
However, it needs to be emphasized that \ac{EA} architectures need to be specifically %
designed to handle dynamically \mbox{(de-)allocated} resources during runtime. %
Therefore, additional research is going on in the area of dynamically scalable %
(often called elastic) software architectures \citep{Brataas:2013:CSM:2479871.2479920}. %

%Due to the aforementioned reasons, %
% Several foundational challenges still need to be solved before such techniques %
% avoid the need for traditional capacity planning and management activities:

Selected challenges in the area of capacity planning and management include %
the following: %

\begin{itemize}
\item Descriptions of the resource demand for \acp{EA} are still too limited in their capabilities (e.g., the amount of resource types they cover).
\item New system architectures such as big data systems require a refocus on storage performance and algorithmic complexity.
\item Energy consumption should be considered as part of a capacity planning activity, as it is a major cost driver in data centers nowadays \citep{Poess:2008:ECK:1454159.1454162}.
\item The use of existing capacity planning and management approaches needs to be simplified to avoid the need to have highly-skilled performance engineers on board at the time of planning a deployment as this is often necessary in a sales process without a project team \citep{Grinshpan:2012:SEA:2207820}.
\end{itemize}
\subsection{Software Architecture Optimization for Performance}
\label{sec:architecturalImprovements}

Whenever a change triggered a (re)design phase, there is also an opportunity to question the current architecture model and find potential for improvement. Architecture-based performance prediction is not limited to design decisions that are directly affected by incoming changes. Additionally, other design decisions taken can be reconsidered in the light of the changed requirements and/or changed environment. As a key aspect of the overall DevOps culture is automation, the remainder of this section will discuss existing tools that can help to automatically achieve such improvements.

A number of approaches that automatically derive performance-optimized software architectures have been surveyed by \citet{aleti2013a}. Up to 2011, the authors found 51 approaches that aim to optimize some performance property. These approaches usually focus on specific types of changes. For example, 37 of the approaches studied allocation of components, 6 approaches address component selection, and 20 approaches address service selection. Overall, the explored changes were allocation, hardware replication, hardware selection, 
software replication, scheduling, component selection, service selection, software selection, service composition, software parameters, clustering, 
hardware parameters, architectural pattern, partitioning, or maintenance schedules. Some approaches also considered problem-specific additional changes and 5 approaches were general, i.e., they supported the modeling of any type of change. 

In addition to optimizing performance, most of the approaches also take potentially conflicting additional objectives into account. Most commonly, costs of the solution are considered as well (38 approaches), followed by reliability (25), availability (18), and energy (6).

Performance models extracted from production systems as outlined in \autoref{sec:modelsatruntime} can be used as a basis to optimize performance. For formulating an \textit{optimization problem} on an architectural performance model, it is required to specify an \textit{objective function} and a list of \textit{possible changes} to be explored by the optimization algorithm. 

\paragraph*{Objective Function}
The associated solvers of the performance model already provide possible evaluation functions. Such an evaluation function takes an architectural performance model as an input and determines performance metrics, e.g., the mean response time, as an output. Selecting a performance metric of interest, such as mean response time, we can easily define an objective function for an optimization problem. Let $M$ denote the set of all valid architectural performance models for a given architecture metamodel (e.g., all valid instances of the Palladio Component Model). Let $f_p: M \rightarrow \mathbb{R}$ denote the evaluation function that determines the selected performance metric $p$ for a given model $m \in M$. Then, $f_p$ can serve as an objective function to optimize architecture models. In case of mean response time $mrt$, we aim to minimize $f_{mrt}$. 

\paragraph*{Possible Changes}
In addition to defining what the objective function is, the other major component of the optimization is defining what can be changed. When optimizing architectural models for performance, we usually want to keep the functionality of the system unchanged. Thus, we do not want to arbitrarily change the model, but only explore functionally-equivalent models with varying performance properties. For example, in component-based architectures, the deployment of components to servers can change without changing the functionality of the system. Likewise, assigning a cache component or replacing middleware components should not change functionality.

There are different approaches on how to encode possible changes. One is to enumerate the set of open design decisions \citep{Koziolek2013}. For example, a decision could be on which server to allocate a component. Another example for an open decision could be whether to add a cache and where. Each open decision has a set of possible alternative options. For example, the possible options for the first design decision could be that a component can be allocated to a set of servers. As another example, the possible options for the second open design decision could be that the cache can be placed in front of a set of components or nowhere. 

Then, a single architectural candidate can be characterized by which option has been chosen for each open decision. One can picture this as a multidimensional decision space where each open decision is a dimension and each possible architectural candidate is a point in this space. 
In addition, it may be required to specify additional constraints on the decision space, such as that component A and component B may not be allocated to the same machine, e.g., due to security concerns. Thus, some of the architectural candidates in the decision space may be invalid. 

\paragraph*{Optimization Problem}
Then, combined, we can define an optimization problem\index{Optimization problem}. A single-objective optimization could be to minimize the objective function $f_{p}$ for the chosen performance metric of interest $p$ over the valid architectural candidates in the decision space. If multiple objective functions are of interest, one can also formulate a multi-objective problem with several objective functions $f_{p_1}$, ..., $f_{p_n}$ and search for the so-called Pareto-optimal solutions \citep{Deb2005}. A solution is Pareto-optimal, if one cannot find another solution that is better or at least equal with respect to all objective functions.\\

%%% Challenges
Even though a lot of approaches exist to automatically improve a software architecture for performance and it is known how to specify a general optimization problem based on performance models, a few major challenges remain:
\begin{itemize}
  \item All the approaches that use performance models as input for a software architecture optimization rely on the accuracy of the information represented in the model. Whenever a certain aspect of a software system is not represented, it cannot be optimized. It thus may be necessary to derive different model granularities for runtime optimization of systems and general architecture optimization.
  \item Even though automatic performance model generation approaches exist, the specification of the possible changes to these models remains a manual step. It remains to be seen how the specification of these possibilities can be designed to make it simple for the users and thus increase the adoption rate.
  \item Most software architecture optimization approaches surveyed by \citet{aleti2013a} are limited in that they either focus on specific possible changes only, that they only support simple performance prediction (e.g., very simple queuing models), or that they consider no or few conflicting objectives. Thus, a general optimization framework for software architectures could be devised, which could make use of %
  \begin{inlineenumerate}
  \item plug-ins that interpret different architecture models (from architecture description languages to component models) and provide degree of freedom definitions and %
  \item plug-ins to evaluate quality attributes for a given architecture model. 
  \end{inlineenumerate}

\end{itemize}

\section{Conclusion}
\label{sec:conclusion}

This report outlined activities assisting the performance-oriented DevOps integration with the help of measurement- and model-based performance evaluation techniques. The report explained performance management activities in the whole life cycle of a software system and presented corresponding tools and studies. Following a general section about existing measurement- and model-based performance evaluation techniques, the report focused on specific activities in the development and operation phase. Afterwards, it outlined activities during the evaluation phase when a system is going back-and-forth between development and operation.

A key success factor for all the integrated activities outlined in this report is the interoperability between the different tools and techniques. For example, an architect might create a deployment architecture of a software system, conduct several studies, then move the model to a tool better suited to performance analysis. For models, approaches such as the \ac{PMIF}  \citep{Smith2010} exist to help in this process. When someone from operations might want to  communicate metrics to someone from development, it is necessary to be able to exchange the metrics in a common format. For this use case, formats such as the \ac{CIM} Metrics Model \citep{CIM2003MM}, the \ac{SMM} \citep{OMG2012SMM} or the performance monitoring specification of the %\ac{OSLC}
\citet{OSLC2014PMS} exist. However, even though approaches exist, they need to be supported by multiple vendors in order for them to 
work. It is still to be seen which of these approaches and specifications might establish themselves as an industry standard.

As of today the outlined approaches exist in theory and practice, but most model-based approaches are developed in academia and not in industry context. Furthermore, most development and operations activities are not tightly integrated as of today. Integrating the proposed approaches and increasing the degree of automation are key challenges for applying performance models in industry context and supporting \ac{Dev} and \ac{Ops} in terms of performance improvements. Several approaches focus on specific systems and need to be generalized for broader usage scenarios. Further validation on large industry projects would increase the level of trust and readiness to assume the costs and risks of applying performance models. Both, industry and academia have to address these challenges to enable a fully performance-oriented DevOps integration.

\cleardoublepage

\newpage
\section{Acronyms}
\begin{acronym}[UML-SPT]
\acro{APM}{Application Performance Management}
\acro{API}{Application Program Interface}
\acro{CBMG}{Customer Behavior Model Graph}
\acro{CD}{Continuous Delivery}
\acro{CDE}{Continuous Deployment}
\acro{CI}{Continuous Integration}
\acro{CIM}{Common Information Model}
\acro{CPU}{Core Processing Unit}
\acro{CSM}{Core Scenario Model}
\acro{Dev}{development}
\acro{DLIM}{Descartes Load Intensity Model}
\acro{DML}{Descartes Modeling Language}
\acro{EA}{enterprise application}
\acro{EFSM}{Extended Finite State Machine}
\acro{EJB}{Enterprise Java Bean}
% \acro{ERP}{Enterprise Resource Planning}
\acro{EUM}{end-user experience monitoring}
\acro{HDD}{hard disk drive}
\acro{Java EE}{Java Enterprise Edition}
\acro{JPA}{Java Persistence API}
\acro{IDE}{Integrated Development Environment}
\acro{I/O}{Input / Output}
\acro{IT}{Information Technology}
\acro{ITOA}{IT operations analytics}
\acro{LibReDE}{Library for Resource Demand Estimation}
\acro{LIMBO}{Load Intensity Modeling Tool}
\acro{LQN}{Layered Queueing Network}
\acro{MARTE}{\acs{UML} Profile for Modeling and Analysis of Real-Time and Embedded Systems}
\acro{PCM}{Palladio Component Model}
\acro{PMIF}{Performance Model Interchange Format}
\acro{QN}{Queueing Network}
\acro{QPN}{Queueing Petri Net}
\acro{OSLC}{Open Services for Lifecycle Collaboration}
\acro{Ops}{operations}
\acrodefplural{Ops}[Ops]{operations}
\acro{SDL}{Specification and Description Language}
\acro{SLA}{Service-level Agreement}
\acro{SMM}{Structure Metrics Meta-Model}
\acro{SPE}{Software Performance Engineering}
\acro{SPL}{Stochastic Performance Logic}
\acro{SUA}{System Under Analysis}
% \acro{SUT}{System Under Test}
% \acro{TCO}{Total Cost of Ownership}
\acro{UI}{user interface}
\acro{UML}{Unified Modeling Language}
\acro{UML-SPT}{\acs{UML} Profile for Schedulability, Performance and Time}
\end{acronym}

%% --------------------
%% |   Bibliography   |
%% --------------------
\cleardoublepage
%\phantomsection
\renewcommand\bibname{References}
\addcontentsline{toc}{section}{\bibname}

\bibliographystyle{abbrnat2}

\bibliography{maindoc}

\end{document}